\documentclass[structabstract]{aa}  
\usepackage{graphicx}
\usepackage{natbib}
\usepackage{longtable}
\usepackage{txfonts}

\begin{document}

   \title{HST/ACS colour-magnitude diagrams of M31 globular clusters \thanks{Based 
   on observations made with the NASA/ESA {\em Hubble Space Telescope}, obtained 
   from the data archive at the Space Telescope Science Institute. STScI is operated
    by the Association of Universities for Research in Astronomy, Inc., under NASA 
    contract NAS 5-26555."}}
   \author{S. Perina\inst{1}, L. Federici\inst{2}, M. Bellazzini\inst{2}, C. Cacciari\inst{2}, 
F. Fusi Pecci\inst{2},\and S. Galleti\inst{2} }

   \offprints{S. Perina}

   \institute{Universit\`a di Bologna, Dipartimento di Astronomia
             Via Ranzani 1, 40127 Bologna, Italy\\
            \email{sibilla.perina@unibo.it}
         \and
             INAF - Osservatorio Astronomico di Bologna,
              Via Ranzani 1, 40127 Bologna, Italy\\
          \email{luciana.federici@oabo.inaf.it,michele.bellazzini@oabo.inaf.it,  \\
          flavio.fusipecci@oabo.inaf.it, carla.cacciari@oabo.inaf.it, silvia.galleti@oabo.inaf.it }
      }
     \authorrunning{S. Perina et al.}

   \date{Received May 28, 2009; accepted September 23, 2009}

\abstract
{}
{With the aim of increasing the sample of M31 clusters for which a colour-magnitude
 diagram is available, we searched the HST archive for ACS images 
containing objects included in the Revised Bologna Catalogue of M31 
globular clusters \thanks{RBC Version 3.5 available at: www.bo.astro.it/M31}.}
{Sixty-three such objects were found. We used the ACS images to confirm or revise their
classification and were able to obtain useful CMDs for 11  
old globular clusters and 6 luminous young clusters. We obtained simultaneous
estimates of the distance, reddening, and metallicity of old clusters by
comparing their observed field-decontaminated CMDs with a grid of template
clusters of the Milky Way. We estimated the age of the young clusters by fitting
with theoretical isochrones.}
{For the old clusters, we found metallicities in the range $-0.4\le $[Fe/H]$\le
-1.9$. The individual estimates generally agree with existing
spectroscopic estimates. At least four of them display a clear blue horizontal
branch, indicating ages $\ga 10$ Gyr. All six candidate young clusters are
found to have ages $<1$ Gyr. The photometry of the clusters is made publicly
available through a dedicated web page.}
{With the present work the total number of M31 GCs with reliable optical CMD increases from
35 to 44 for the old clusters, and from 7 to 11 for the young ones.
The old clusters show similar characteristics to those of the MW.
We discuss the case of the cluster B407, with a metallicity [Fe/H]$\simeq -0.6$
and located at a large projected distance from the centre of M31 (R$_p=19.8$
kpc) and from the major axis of the galaxy (Y$=11.3$ kpc). Metal-rich globulars
at large galactocentric distances are rare both in M31 and in the Milky Way.
B407, in addition, has a velocity in stark contrast with the rotation pattern
shared by the bulk of M31 clusters of similar metallicity. This, along with
other empirical evidence, supports the hypothesis that the cluster (together 
with B403) is physically associated with a substructure in the halo of M31 that 
has been interpreted as the relic of a merging event. }  

   \keywords{Galaxies:individual:M31 -- Galaxies:star clusters -- Catalog
--- (Galaxies:)Local Group}

   \maketitle


\section{Introduction} 

Over the past $\sim$20 years, the globular cluster (GC)  system of M31 has been
the subject of intensive study   both from the ground and from space-borne
observatories (see Rich et al. 2005; Galleti et al. 2004 - hereafter G04, 2006a, 2007; 
Huxor et al. 2008;  Lee et al. 2008  and  Caldwell et al. 2009 - hereafter C09, 
for recent reviews and references). 
One of the main aims of these studies was to collect as much as
possible information on  the GCs in M31 and compare it with our knowledge of the
GCs in the Galaxy,  so as to derive better insight into the formation and
(chemical and dynamical)  evolution of these two spiral galaxies and possibly of
galaxies in general. The advent of the Hubble Space Telescope provided the unprecedented
opportunity to obtain colour-magnitude  diagrams (CMD) of M31 clusters, thus adding a
completely new perspective to this research.

Substantial contributions in this field have been made by many  investigators. 
At present, sufficiently accurate visual CMDs for a meaningful comparison  with their
Galactic counterparts have been published for 35 GCs in M31. 
Except for one that was observed from the ground (MGC1, Martin et al. 2006), a good fraction of
these have been obtained with the HST-WFPC2 (Ajhar  et  al. 1996,  Rich et al. 1996,
Fusi Pecci et al. 1996, Holland et al. 1997, Jablonka et al.  2000, Meylan et al.
2001, Rich  et al. 2005)  until the better resolution and sensitivity of the ACS
allowed even  more accurate CMDs at fainter limiting magnitudes (Brown
et al. 2004;  Huxor et al. 2004, 2005, 2008; Galleti et al. 2006b; 
Mackey et al. 2006, 2007).

In addition to photometric  quality, which is essential for the analysis  of
individual objects, a good statistical coverage is also important for a better 
understanding of the GC system.  To increase the sample of M31 GCs with a CMD
of individual member stars,  we searched the HST archive for ACS images of
objects that are listed in the Revised Bologna Catalogue of M31 
clusters (RBC, 
see G04). We found useful ACS images containing 69 such objects (see Fig.~\ref{f:m31}). 
The retrieved material allowed us to
confirm or revise the classification of all of them and to obtain CMD of
individual stars for 17, 11 likely old globulars, and 6 young luminous clusters
(like those discussed in Williams \& Hodge 2001; Fusi Pecci et al 2005 and
Perina et al. 2009a). This paper is devoted to the analysis of these data.

In Sect. 2 we present the target list, and in Sect. 3 we describe the adopted 
reduction procedures that yielded the CMDs.   Section 4 is devoted to describe the
method we have used to  estimate the  metallicity, reddening, and distance from
each individual CMD for which a sufficiently reliable decontamination from the
non-member field components  was feasible. In Sect. 5 specific notes and comments
on the results are presented for each  of the 11 GCs (the primary targets) and for
the other objects for which a sufficiently meaningful photometry was
carried out. In Sect. 6 we discuss a possible connection between 
a few clusters and a large substructure recently found in M31.
Finally, Sect. 7 contains some general considerations and
conclusions. 

\section{The targets}  




\begin{figure*}
 \centering\includegraphics[width=17.6cm]{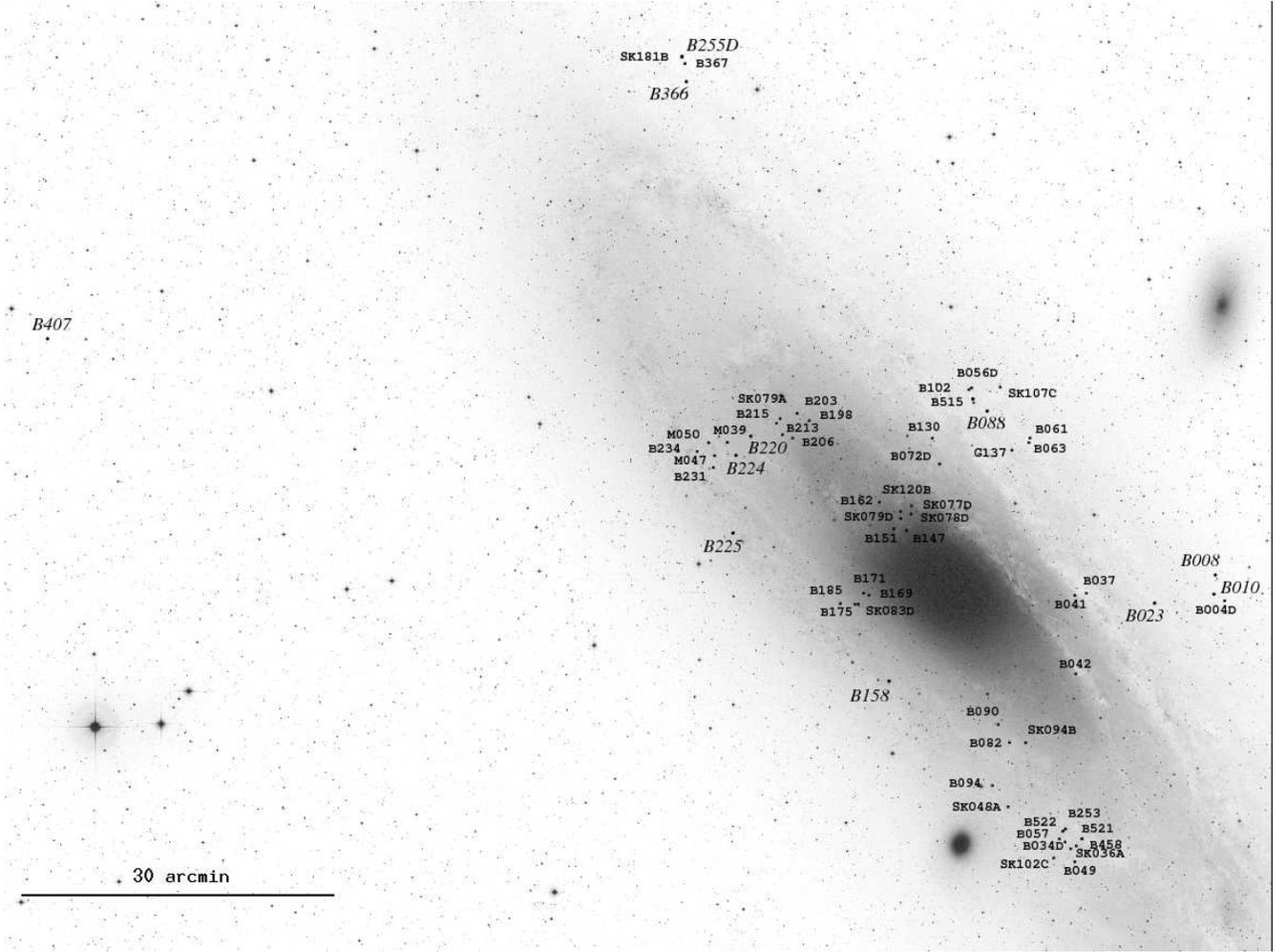}
\caption{The location of the 11 primary target globular clusters, marked 
in italics + 52 secondary targets (see Sect.2 and 5), projected against 
the body of M31, with North up and East to the left.}
\label{f:m31}
\end{figure*}

A search by coordinates allowed us to find ACS images \footnote{released until June 2007 
from the HST Archive.} for 69 entries of
the  RBC V3.5, independently of their
original classification (see G04, and Galleti et al. 2006a). In two
cases the images revealed that there were two catalogue entries referring to the
same object (i.e. B521=SK034A, and B522=SK038A), thus reducing the number of real
objects to 67. Four confirmed clusters classified as candidate "intermediate-age
GCs" by Puzia et al. (2005), and for which we have obtained good CMDs, have been
excluded from the list as they will be the subject of a dedicated study (Perina et
al. 2009b, in preparation). 

Eighteen of the remaining 63 objects, namely  B004D, B253, B034D, SK102C,
G137, SK107C, B102,  SK094B, B072D, SK077D, SK078D, SK079D,
SK120B, SK083D, B175, SK079A,  M047, and SK181B, are not {\em bona fide} clusters:
their original RBC classification has been confirmed or revised based on the high
resolution ACS images. The results of this analysis are summarised in Table~2
where we report their old and new classification flag. 

Twenty of the remaining 45 objects are unequivocally confirmed as {\em bona
fide} clusters (B037, B041, B042, B056D, B061, B063, B082, B094, SK048A, B130,  B185,
B198, B203, B206, B213, B215, B231, B234, B522=SK038A, and  SK036A, see 
Fig.~\ref{f:field}) and we obtained photometry of individual stars from the 
respective images, however we were unable to find an annulus around the
cluster centre where the population of the cluster could be disentangled from the
population of the surrounding field. In general this is due to the extreme
compactness of the clusters, preventing to obtain good photometry for a sufficient
number of stars even in the outermost coronae, but also the density of the
background population plays a role. For five additional clusters, e.g. B147, B151, 
 B162, B169, B171 
(Fig.~\ref{f:field}), located in the bulge of M31, at projected
distances  R=7.8$\arcmin$,7.29$\arcmin$, 7.17$\arcmin$, 6.31$\arcmin$ and
9.95$\arcmin$ from the centre,  the overall crowding was so high that it resulted
impossible to carry out any meaningful  photometry even in the field, with the
method adopted here.

\begin{figure*}
\centering 
\includegraphics[width=12.cm]{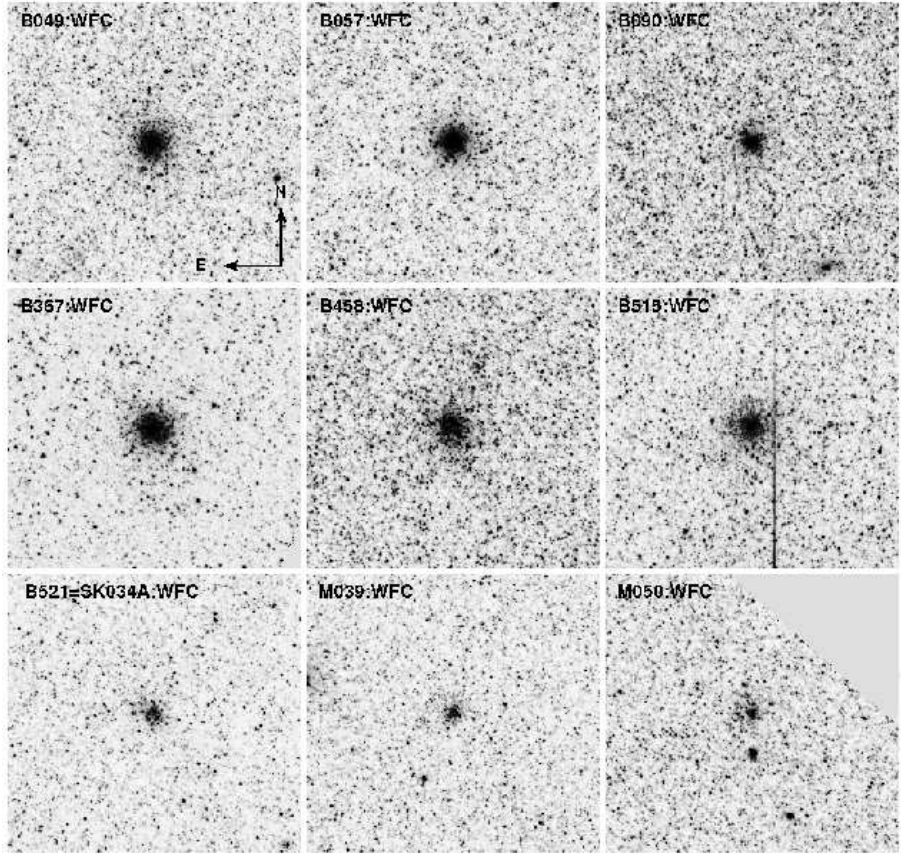}
\caption{ V band (F606W) images of the 9 candidate young clusters (see Sect. 5.12). 
The cluster and ACS camera identification  are shown in each subraster.  
Each image covers  20$^{\arcsec}\times$~20$^{\arcsec}$. 
}
\label{f:blcc}
\end{figure*}

The remaining 20 objects are the main subject of the present analysis and are 
subdivided as follows:

\begin{itemize} 

\item Eleven {\em bona fide} clusters for which we could obtain a meaningful
CMD,  albeit of varying accuracy\footnote{depending on the cluster
characteristics, the  crowding conditions and the surface density of the
surrounding field.},  and that were revealed by their CMD to be likely classical
old globulars (i.e. having ages of several Gyr). These are the "primary
targets" discussed in this  paper, namely B008, B010, B023, B088, B158, B220,
B224, B225, B255D, B366, and B407, according to the RBC nomenclature. 

\item Nine {\em bona fide} clusters that were listed as candidate young
clusters  (age $\la 2$ Gyr) by some previous study (Fig.~\ref{f:blcc}) .  Five
of them, namely B049, B057, B090,  B367, B458 were included in the list of the
so-called "Blue Luminous Compact Clusters"  (Fusi Pecci et al. 2005);  three of
them, namely B521=SK034A, M039=KHM31-516 (Krienke  and Hodge 2008), and M050
were classified as "young" by C09 (see Table \ref{t:young});
and one, B515=KHM31-409, was included in the list of possible young/open
clusters of Krienke and Hodge (2008). For six of them (B039, B049, M050,  B367,
B458, and B521) we were able to derive a CMD in which the cluster population
can be identified and we can confirm their young age, while for the other three
we obtained useful photometry only for the surrounding field.

\end{itemize}

\begin{figure*}
 \centering\includegraphics[width=12.cm]{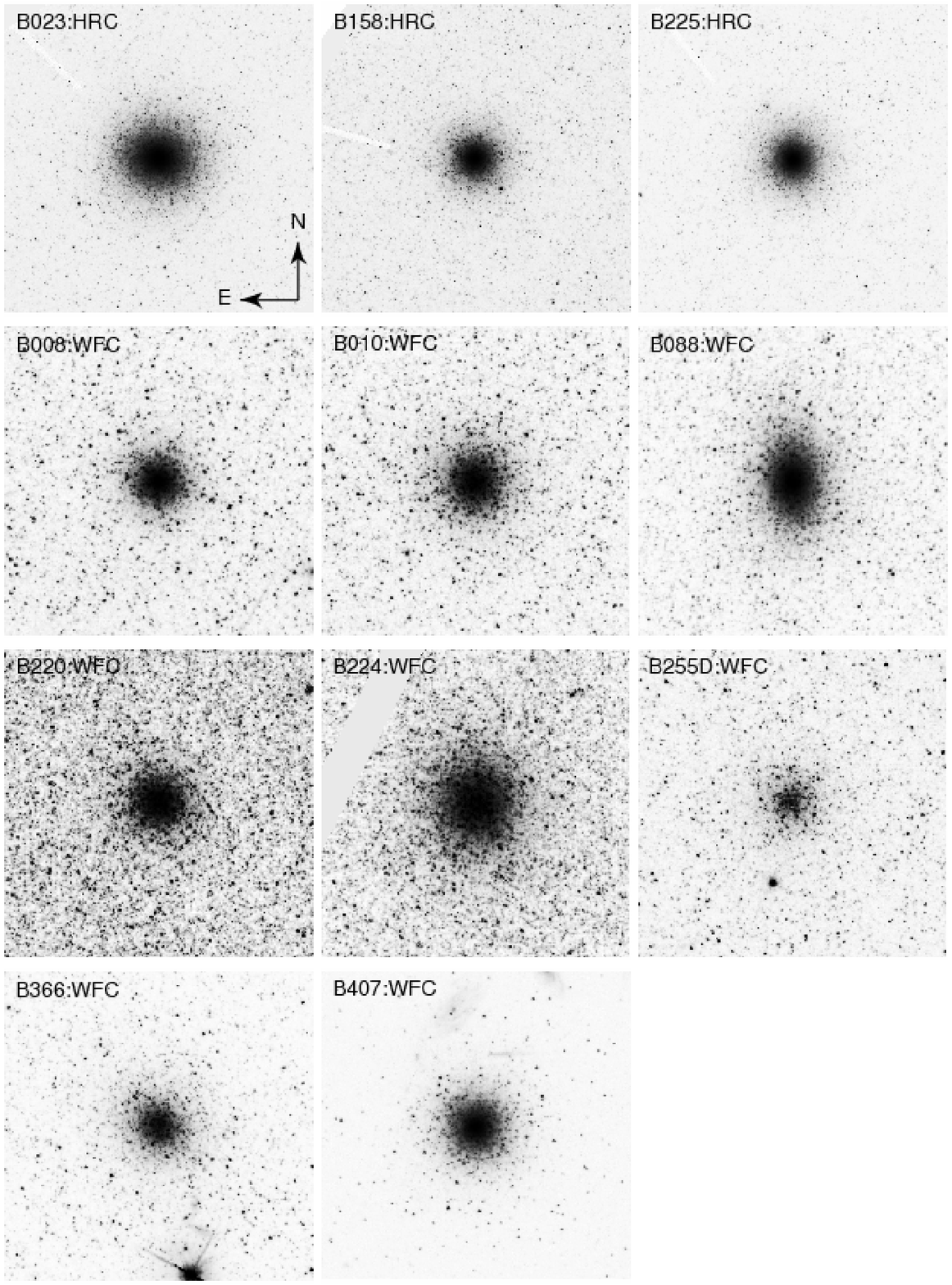}
\caption{ V band (F606W)  images of the 11 M31 GCs  analysed in the present 
study (the primary targets). The cluster and ACS camera identification  
are shown in each subraster.  Each image covers 20$^{\arcsec}\times$~20$^{\arcsec}$ 
(20$^{\arcsec}$ = 76 pc at the assumed M31 distance modulus of 24.47). 
North is up and East to the left.
}
\label{f:gc}
\end{figure*}

\begin{figure*}
\centering
\includegraphics[width=15.cm]{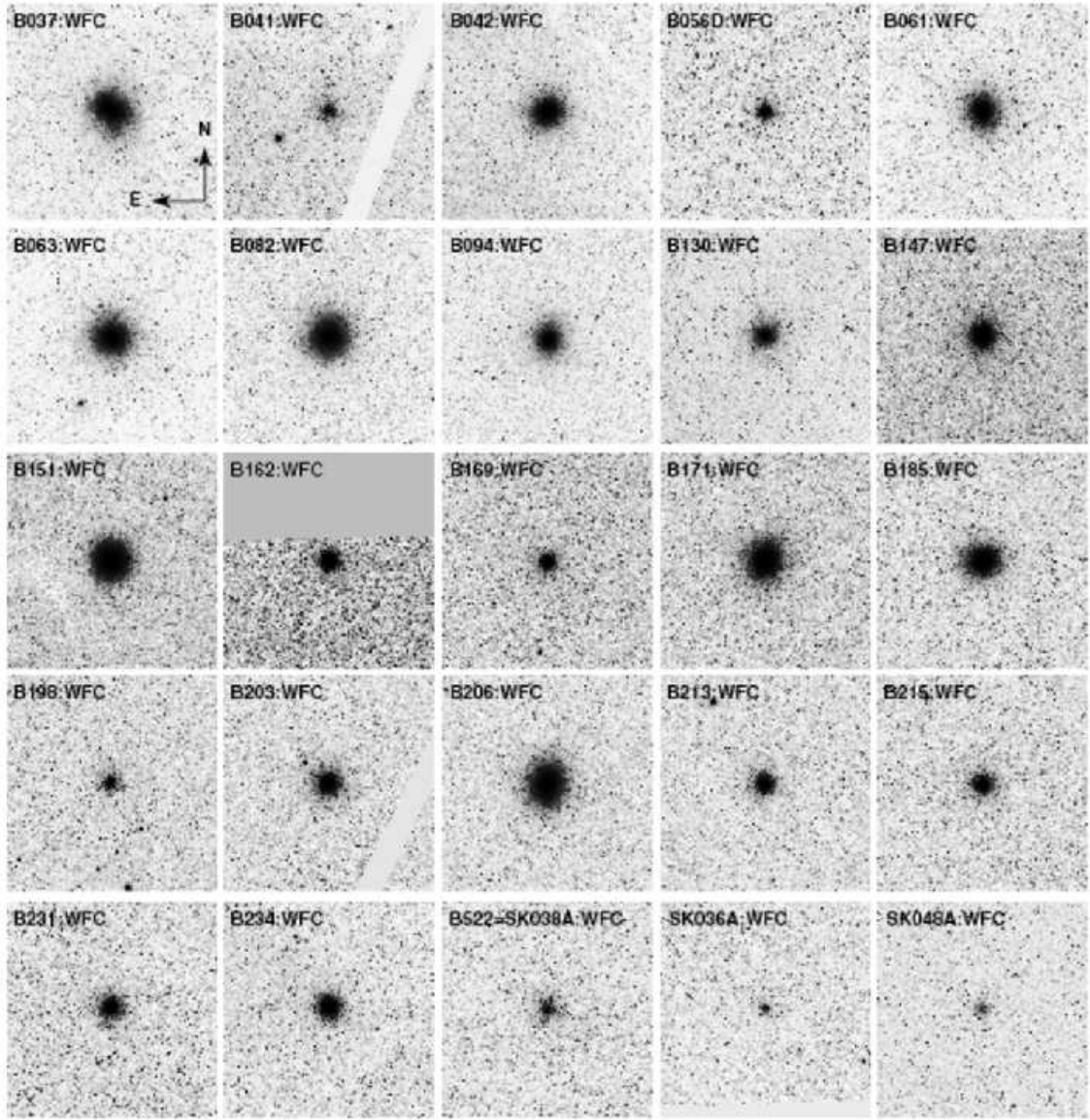}
\caption{Same as in Fig. \ref{f:gc} for 25 additional M31 globular clusters (see Sect. 2 ).
}
\label{f:field}
\end{figure*}

Going back to the 11 ``primary target'' GCs discussed in detail in the present 
study, most of them lie close to the galactic plane of M31, as shown  in Fig.
\ref{f:m31}.  Three of them have been observed with the ACS/HRC and eight with
the ACS/WFC. Their V images are shown in Fig. \ref{f:gc} and their HST data are
listed in  Table \ref{t:gc}, together with their integrated magnitudes and
colours taken from the  RBC, when available. Similar data for all the other 52 
targets considered in this paper are reported in Table  \ref{t:gcnocmd}.


\begin{table*} 
 \centering
 \caption{ The primary target M31 GCs. ID, coordinates and photometry are from G04;
 [Fe/H] are from ($^a$):  Perrett et al. (2002), and ($^b$): Huchra, Brodie \& Kent(1991).}
\label{t:gc}
  \begin{tabular}{@{}lc@{\hspace{0.05in}}c@{\hspace{0.1in}}r@{\hspace{0.05in}}rc@{\hspace{0.05in}}c@{\hspace{0.045in}}c@{\hspace{0.1in}}clr@{}}
  \hline
\\
{ }ID  & RA (J2000) & Dec (J2000) &\multicolumn{1}{c}{X}&\multicolumn{1}{c}{Y}& V  & (B-V) & (V-I) & [Fe/H] &ACS camera, bands(total $t_{exp}$) & PID{ }  \\
       &            &             &\multicolumn{1}{c}{arcm} &\multicolumn{1}{c}{arcm} &    &       &       &        &                         &         \\
 \hline
\\
  B008-G60   & 00 40 30.54 & +41 16 09.7 & --15.41 &   19.86 & 16.56 &  1.10 &  1.05 & --0.41$^a$ & WFC,F606W(3250s),F435W(7260s) & 10407 \\  
  B010-G62   & 00 40 31.56 & +41 14 22.3 & --16.71 &   18.62 & 16.66 &  0.84 &  1.18 & --1.87$^b$ & WFC,F606W(3250s),F435W(7260s) & 10407  \\  
  B023-G78   & 00 41 01.26 & +41 13 45.3 & --13.78 &   13.82 & 14.22 &  1.18 &  1.65 & --0.92$^b$ & HRC,F606W(2020s),F814W(2860s) &  9719  \\ 
  B088-G150  & 00 42 21.10 & +41 32 14.3 &   10.00 &   13.32 & 15.42 &  1.12 &  1.47 & --2.17$^b$ & WFC,F606W(2370s),F814W(2370s) & 10260   \\  
  B158-G213  & 00 43 14.47 & +41 07 20.6 &  --3.45 &  --9.90 & 14.70 &  0.86 &  1.15 & --1.08$^b$ & HRC,F606W(2020s),F814W(2860s) &  9719  \\
  B220-G275  & 00 44 19.49 & +41 30 35.7 &   22.38 &  --5.13 & 16.55 &  0.78 &  1.06 & --2.07$^b$ & WFC,F606W(1860s),F435W(2910s) & 10407   \\                
  B224-G279  & 00 44 27.21 & +41 28 50.6 &   21.90 &  --7.35 & 15.45 &  0.79 &  1.03 & --1.90$^b$ & WFC,F606W(1860s),F435W(2910s) & 10407  \\                
  B225-G280  & 00 44 29.78 & +41 21 36.6 &   16.52 & --12.21 & 14.15 &  1.01 &  1.39 & --0.70$^b$ & HRC,F606W(2020s),F814W(2860s) &  9719  \\
  B255D-D072 & 00 44 48.55 & +42 06 13.3 &   53.70 &   12.71 & 17.92 &       &       &            & WFC,F606W(1850s),F435W(2920s) & 10407  \\                     
  B366-G291  & 00 44 46.72 & +42 03 50.3 &   51.62 &   11.49 & 15.99 &  0.81 &  1.01 & --1.39$^b$ & WFC,F606W(1850s),F435W(2920s) & 10407   \\                
  B407-G352  & 00 50 09.98 & +41 41 01.1 &   71.54 & --49.72 & 16.05 &  0.90 &  1.22 & --0.85$^b$ & WFC,F606W(2400s),F814W(5100s) &  9458   \\
\\
 \hline
\end{tabular} 
\end{table*} 



\begin{table*} 
\centering
\caption{The additional targets (see Sect. 2) grouped according to their location within the same-exposure field.
All of them were observed with WFC@ACS on HST. ID, coordinates and photometry are from Galleti et 
al. (2004). We note the double identifications B521=SK034A and B522=SK038A.}
\scriptsize
\label{t:gcnocmd}
  \begin{tabular}{@{}lc@{\hspace{0.05in}}c@{\hspace{0.1in}}r@{\hspace{0.05in}}rc@{\hspace{0.05in}}c@{\hspace{0.045in}}c@{\hspace{0.1in}}c@{\hspace{0.08in}}l@{\hspace{0.05in}}lcl@{}}  
  \hline
\\
ID  & RA (J2000) & Dec (J2000) &\multicolumn{1}{c}{X}&\multicolumn{1}{c}{Y}& V  & (B-V) & (V-I) & [Fe/H] &type$^\ast$&\multicolumn{1}{c}{bands(exptime)} & PID{ }&{\hspace{0.25in}}Datasets \\ 
       &            &             &\multicolumn{1}{c}{arcm} &\multicolumn{1}{c}{arcm} &    &       &     &       &        &       &       &               \\                 
 \hline                                                                                                                                                   
\\
  B004D-V223          & 00 40 26.41 & +41 13 42.7 & --17.82 &  18.98 & 18.81 & 1.18 &      &             &4& F606W(3250s),F435W(7260s) & 10407& J96Q07010,J96Q07020 \\  
&  &  & & & &  &  &  & &  &\\  
  B037-V327               & 00 41 35.00 & +41 14 54.9 &  --8.98 &   9.51 & 16.82 & 2.05 & 2.63 &  --1.07$^a$ &1& F606W(2370s),F814W(2370s)    & 10260& J8Z003010,J8Z003020 \\  
  B041-G103               & 00 41 40.73 & +41 14 45.8 &  --8.44 &   8.57 & 17.65 & 0.97 & 1.18 &  --1.22$^a$ &1& F606W(2370s),F814W(2370s)    & 10260& J8Z019010,J8Z019010 \\  
&  &  & & & &  &  &  & &  &  &\\  
  B042-G104               & 00 41 41.69 & +41 07 25.8 & --14.12 &   3.93 & 16.29 & 1.48 & 1.89 &  --1.09$^b$ &1& F606W(2370s),F814W(2370s)    & 10260& J8Z060010,J8Z022010 \\  
&  &  & & & &  &  &  & &  &\\  
  B057-G118$^\dagger$    & 00 41 52.84 & +40 52 04.6 & --24.96 & --7.15 & 17.64 & 0.69 & 0.99 &  --2.12$^a$ &1& F606W(2110s),F435W(2672s)    & 10407& J96Q06010,J96Q06020 \\  
  B253                    & 00 41 49.63 & +40 52 59.7 & --24.60 & --6.11 & 18.01 &      & 0.55 &             &6& F606W(2110s),F435W(2672s)    & 10407& J96Q06010,J96Q06020 \\  
  B034D                   & 00 41 50.13 & +40 51 46.7 & --25.51 & --6.93 & 17.50 &      &      &             &6& F606W(2110s),F435W(2672s)    & 10407& J96Q06010,J96Q06020 \\  
  B522-SK038A             & 00 41 50.94 & +40 52 48.3 & --24.60 & --6.42 & 17.85 &      &      &             &1(2)& F606W(2110s),F435W(2672s) & 10407& J96Q06010,J96Q06020  \\  
  SK102C                  & 00 41 55.92 & +40 50 19.7 & --25.98 & --8.68 & 15.22 & 0.76 & 0.85 &             &6(2)& F606W(2110s),F435W(2672s) & 10407& J96Q06010,J96Q06020  \\  
  B521-SK034A             & 00 41 41.67 & +40 52 01.4 & --26.29 & --5.51 &       &      &      &             &1(2)$^\ddagger$& F606W(2110s),F435W(2672s) & 10407& J96Q06010,J96Q06020  \\  
  B458-D049$^\dagger$    & 00 41 44.61 & +40 51 22.3 & --26.47 & --6.35 & 17.84 & 0.49 & 0.57 &  --1.18$^a$ &1& F606W(2110s),F435W(2672s)    & 10407& J96Q06010,J96Q06020  \\  
  B049-G112$^\dagger$    & 00 41 45.60 & +40 49 53.7 & --27.52 & --7.41 & 17.56 & 0.52 & 0.69 &  --2.14$^a$ &1& F606W(2110s),F435W(2672s)    & 10407& J96Q06010,J96Q06020  \\  
  SK036A                  & 00 41 47.34 & +40 51 07.5 & --26.35 & --6.91 & 19.43 & 1.01 & 1.13 &             &1& F606W(2110s),F435W(2672s)    & 10407& J96Q06010,J96Q06020  \\  
&  &  & & & &  &  &  & &  &  &\\  
  B063-G124               & 00 42 00.80 & +41 29 09.5 &    5.24 &  14.43 & 15.66 & 1.21 & 1.58 &  --0.87$^b$ &1& F606W(2370s),F814W(2370s)    & 10260& J8Z008010,J8Z024010  \\  
  B061-G122               & 00 42 00.20 & +41 29 35.5 &    5.51 &  14.79 & 16.61 & 1.12 & 1.49 &  --0.79$^b$ &1& F606W(2370s),F814W(2370s)    & 10260& J8Z008010,J8Z024010  \\  
  G137                    & 00 42 09.43 & +41 28 30.4 &    5.71 &  12.76 & 17.81 & --0.02 &    &             &5& F606W(2370s),F814W(2370s)    & 10260& J8Z008010,J8Z024010  \\  
&  &  & & & &  &  &  & &  &  &\\  
  SK107C              & 00 42 14.18 & +41 34 26.3 &   10.94 &  15.70 & 19.65 & 0.80 & 0.89 &             &6(2)& F606W(2370s),F814W(2370s) & 10260& J8Z007010,J8Z023010  \\  
  B515$^{1}$                & 00 42 28.05 & +41 33 24.5 &   11.72 &  13.02 & 18.67$^{2}$ &      &      &         &1& F606W(2370s),F814W(2370s)    & 10260& J8Z007010,J8Z023010  \\  
  B056D               & 00 42 28.45 & +41 34 27.2 &   12.59 &  13.60 & 18.70 &      &      &             &1& F606W(2370s),F814W(2370s)    & 10260& J8Z007010,J8Z023010  \\  
  B102                & 00 42 29.85 & +41 34 18.2 &   12.64 &  13.30 & 16.58 & 0.62 & 0.95 &             &7& F606W(2370s),F814W(2370s)    & 10260& J8Z007010,J8Z023010  \\  
&  &  & & & &  &  &  & &  &  &\\  
  B082-G144               & 00 42 15.79 & +41 01 14.3 & --15.06 & --4.94 & 15.54 & 1.56 & 1.91 &  --0.86$^b$ &1& F606W(2370s),F814W(2370s)    & 10260& J8Z004010,J8Z020010  \\  
  SK094B                  & 00 42 07.81 & +41 01 10.0 & --16.05 & --3.80 & 18.13 & 1.11 & 1.24 &  --0.86$^b$ &4(2)& F606W(2370s),F814W(2370s) & 10260& J8Z004010,J8Z020010  \\  
  B090$^\dagger$         & 00 42 21.12 & +41 02 57.3 & --13.09 & --4.68 & 18.80 &      & 1.64 &  --1.39$^a$ &1& F606W(2370s),F814W(2370s)    & 10260& J8Z004010,J8Z020010  \\  
&  &  & & & &  &  &  & &  &  &\\  
  B094-G156               & 00 42 25.01 & +40 57 17.2 & --17.11 & --8.74 & 15.55 & 0.97 & 1.26 &  --0.41$^b$ &1& F555W( 413s),F814W(502s)     & 10273& J92GB9BRQ,J92GB9BPQ \\  
  SK048A                  & 00 42 17.59 & +40 55 15.3 & --19.58 & --8.89 & 18.49 & 0.65 & 0.74 &             &1& F555W( 413s),F814W(502s)     & 10273& J92GB9BRQ,J92GB9BPQ  \\  
&  &  & & & &  &  &  & &  &  &\\  
  B130-G188               & 00 42 48.91 & +41 29 52.9 &   11.35 &   7.77 & 16.93 & 1.15 & 1.41 &  --1.28$^a$ &1& F555W( 413s),F814W(502s)     & 10273& J92GB6ZLQ,J92GB6ZNQ  \\  
  B072D                   & 00 42 45.78 & +41 27 26.9 &    9.07 &   6.74 & 18.50 &      &      &  --1.28$^a$ &3(4)$^3$ & F555W( 413s),F814W(502s)  & 10273& J92GB6ZLQ,J92GB6ZNQ  \\  
&  &  & & & &  &  &  & &  &  &\\  
  B151-G205               & 00 43 09.64 & +41 21 32.6 &    7.17 & --0.43 & 14.83 & 1.23 & 1.45 &  --0.75$^b$ &1& F606W(2370s),F814W(2370s)    & 10260& J8Z005010,J8Z021010  \\  
  B147-G199               & 00 43 03.31 & +41 21 21.5 &    6.30 &   0.39 & 15.80 & 0.84 & 1.27 &  --0.24$^b$ &1& F606W(2370s),F814W(2370s)    & 10260& J8Z005010,J8Z021010  \\  
  SK077D                  & 00 43 00.52 & +41 23 37.6 &    7.76 &   2.20 & 17.66 & 0.41 & 0.48 &             &6& F606W(2370s),F814W(2370s)    & 10260& J8Z005010,J8Z021010  \\  
  SK078D                  & 00 43 00.86 & +41 22 52.3 &    7.20 &   1.69 & 18.13 & 0.69 & 0.78 &             &6& F606W(2370s),F814W(2370s)    & 10260& J8Z005010,J8Z021010  \\  
  SK079D                  & 00 43 06.04 & +41 22 28.7 &    7.49 &   0.68 & 18.67 & 1.43 & 1.62 &             &6& F606W(2370s),F814W(2370s)    & 10260& J8Z005010,J8Z021010  \\  
  SK120B                  & 00 43 05.97 & +41 23 08.1 &    8.00 &   1.10 & 19.33 & 0.82 & 0.92 &             &6(2)& F606W(2370s),F814W(2370s) & 10260& J8Z005010,J8Z021010  \\  
  B162-G216               & 00 43 16.42 & +41 24 04.2 &    9.95 &   0.13 & 17.48 & 1.05 & 1.34 &             &1& F606W(2370s),F814W(2370s)    & 10260& J8Z005010,J8Z021010  \\  
&  &  & & & &  &  &  & &  &  &\\  
  B171-G222               & 00 43 25.67 & +41 15 37.4 &   4.37  & --6.45 & 15.28 & 0.99 & 1.58 &  --0.48$^b$ &1& F606W(3396s),F435W(4476s)    & 10407& J96Q03010,J96Q03020  \\  
  B169                    & 00 43 23.06 & +41 15 25.5 &   3.91  & --6.19 & 17.08 & 1.23 & 1.31 &             &1& F606W(3396s),F435W(4476s)    & 10407& J96Q03010,J96Q03020  \\  
  SK083D                  & 00 43 28.60 & +41 14 36.7 &   3.92  & --7.51 & 14.64 & 1.05 & 1.17 &             &6& F606W(3396s),F435W(4476s)    & 10407& J96Q03010,J96Q03020  \\  
  B175                    & 00 43 30.18 & +41 14 36.4 &   4.09  & --7.74 & 16.80 & 0.80 &      &             &6& F606W(3396s),F435W(4476s)    & 10407& J96Q03010,J96Q03020  \\  
  B185-G235               & 00 43 37.41 & +41 14 43.3 &   5.02  & --8.74 & 15.54 & 0.94 & 1.18 &  --1.03$^b$ &1& F606W(3396s),F435W(4476s)    & 10407& J96Q03010,J96Q03020  \\  
&  &  & & & &  &  &  & &  &  &\\  
  B206-G257               & 00 43 58.70 & +41 30 18.0 &   19.74 & --2.25 & 15.06 & 0.80 & 1.03 &  --1.45$^b$ &1& F606W(2110s),F435W(2672s)    & 10407& J96Q05010,J96Q05020  \\  
  B198-G249               & 00 43 50.07 & +41 31 53.1 &   19.99 & --0.00 & 17.55 & 0.60 & 1.11 &  --1.13$^a$ &1& F606W(2110s),F435W(2672s)    & 10407& J96Q05010,J96Q05020  \\  
  B203-G252               & 00 43 56.00 & +41 32 36.0 &   21.23 & --0.43 & 16.68 & 0.93 & 1.20 &  --0.90$^a$ &1& F606W(2110s),F435W(2672s)    & 10407& J96Q05010,J96Q05020  \\  
  B213-G264               & 00 44 03.62 & +41 30 38.9 &   20.58 & --2.76 & 16.78 & 1.05 & 1.29 &  --0.99$^b$ &1& F606W(2110s),F435W(2672s)    & 10407& J96Q05010,J96Q05020  \\  
  SK079A                  & 00 44 04.58 & +41 32 09.3 &   21.88 & --1.97 & 18.63 & 1.11 & 1.23 &             &6(1)& F606W(2110s),F435W(2672s) & 10407& J96Q05010,J96Q05020  \\  
  B215-G266               & 00 44 06.44 & +41 31 43.9 &   21.76 & --2.51 & 17.13 & 1.02 & 1.20 &             &1& F606W(2110s),F435W(2672s)    & 10407& J96Q05010,J96Q05020  \\  
&  &  & & & &  &  &  & &  &  &\\  
  M039$^4$                & 00 44 31.30 & +41 30 04.6 &   23.34 & --7.18 & 18.94 & 1.11 &--0.53&             &1(2)$^\ddagger$& F606W(1860s),F435W(2910s) & 10407& J96Q02010,J96Q02020  \\  
&  &  & & & &  &  &  & &  &  &\\  
  B234-G290               & 00 44 46.50 & +41 29 18.3 &  24.50  & --9.90 & 16.78 & 1.00 & 1.18 &  --0.95$^a$ &1& F606W(3315s),F435W(4560s)    & 10407& J96Q04010,J96Q04020   \\  
  M047                    & 00 44 37.85 & +41 28 52.1 &  23.16  & --8.90 & 18.84 &      &  1.2 &             &2$^5$ & F435W(4560s),F606W(3315s) & 10407& J96Q04010,J96Q04020  \\  
  B231-G285               & 00 44 38.61 & +41 27 46.8 &  22.39  & --9.68 & 17.27 & 0.84 & 1.14 &  --1.49$^a$ &1& F435W(4560s),F606W(3315s)    & 10407& J96Q04010,J96Q04020  \\  
  M050                    & 00 44 40.59 & +41 30 06.0 &  24.44  & --8.53 & 18.71 &      & 0.40 &             &1(2)$^\ddagger$& F606W(3315s),F435W(4560s) & 10407& J96Q04010,J96Q04020  \\  
&  &  & & & &  &  &  & &  &  &\\  
  B367-G292$^{\dagger}$ & 00 44 47.18 & +42 05 31.9 &  53.00 &  12.48 & 18.45 & 0.32 & 1.30 &  --2.32$^a$ &1& F606W(1850s),F435W(2920s)    & 10407& J96Q01010,J96Q01020   \\  
  SK181B              & 00 44 48.64 & +42 06 08.1 &   53.64 &  12.64 & 19.18 & 1.28 & 1.46 &             &6(2)& F606W(1850s),F435W(2920s) & 10407& J96Q01010,J96Q01020  \\  
\\
 \hline                                                                               
\end{tabular} 
{\begin{flushleft}{
($^a$): Perrett et al. (2002); ($^b$): Huchra, Brodie \& Kent(1991);  
$^{\dagger}$: BLCC, Fusi Pecci et al. (2005)}
\end{flushleft}}
{\begin{flushleft}{($^{\ast}$): classification, coded as follows: 1- confirmed cluster; 2- gc candidate; 3- controversial object 4- galaxy; 
5- HII region; 6- star; 7- asterism;
 ~~$^\ddagger$: young cluster (from this paper and/or Caldwell et al.(2009)). In parentheses is enclosed the previous RBCv3.5 value. }
\end{flushleft}}
{\begin{flushleft}{($^1$): identified as KHM31-409 in Krienke\&Hodge (2008), tab.4; ~~($^2$): V mag from 
Krienke\&Hodge (2008), tab.4;
~~($^3$): B072D, that was originally classified as a galaxy by Huxor et al. (2008), looks like a cluster, as noted also 
by Caldwell et al. (2009). Radial velocity is necessary  in our view to yield its firm confirmation; 
~~($^4$): identified as KHM31-516 in Krienke \& Hodge (2008), tab.4; ~~($^5$): classified as globular cluster by Caldwell et al. (2009).} 
\end{flushleft}}
\end{table*} 



\section{Data reduction and the colour-magnitude diagrams} 

Data reduction has been performed on the prereduced images provided by STScI, 
using the ACS module of DOLPHOT
\footnote{See http://purcell.as.arizona.edu/dolphot/.} (Dolphin 2000a),
a point-spread function fitting package specifically devoted to the photometry 
of HST data. The package identifies the sources above a fixed flux threshold on
a stacked image and performs the photometry on individual frames, accounts for
the hot-pixel and cosmic-ray masking information attached to the 
observational material, automatically applies
the correction for the Charge Transfer Efficiency (CTE, Dolphin 2000b) 
and transforms instrumental magnitude to the VEGAMAG and standard BVI system 
using the transformations by Sirianni et al. (2005). In the following we use BVI photometry.

We fixed the threshold for the search of sources on the images at 3$\sigma$ above
the background. DOLPHOT provides as output the magnitudes and positions of the
detected sources, as well as a number of quality parameters for a suitable sample
selection, in  view of the actual scientific objective one has in mind. Here we
selected all  the sources having valid magnitude measurements in both passbands,
global  quality flag = 1 (i.e., best measured stars), {\em crowding}  parameter
$\le 0.3$, $\chi^2<1.5$ if V$<22.5$, $\chi^2<2.5$ for brighter stars,
and {\em sharpness} parameter  between -0.3 and 0.3 (see
Dolphin 2000b for details on the parameters). This selection cleans the sample
from the vast majority of spurious and/or bad  measured sources without
significant loss of information, and it has been found  to be appropriate for the
whole data set.

\begin{figure}[t]
 \includegraphics[width=8.5cm]{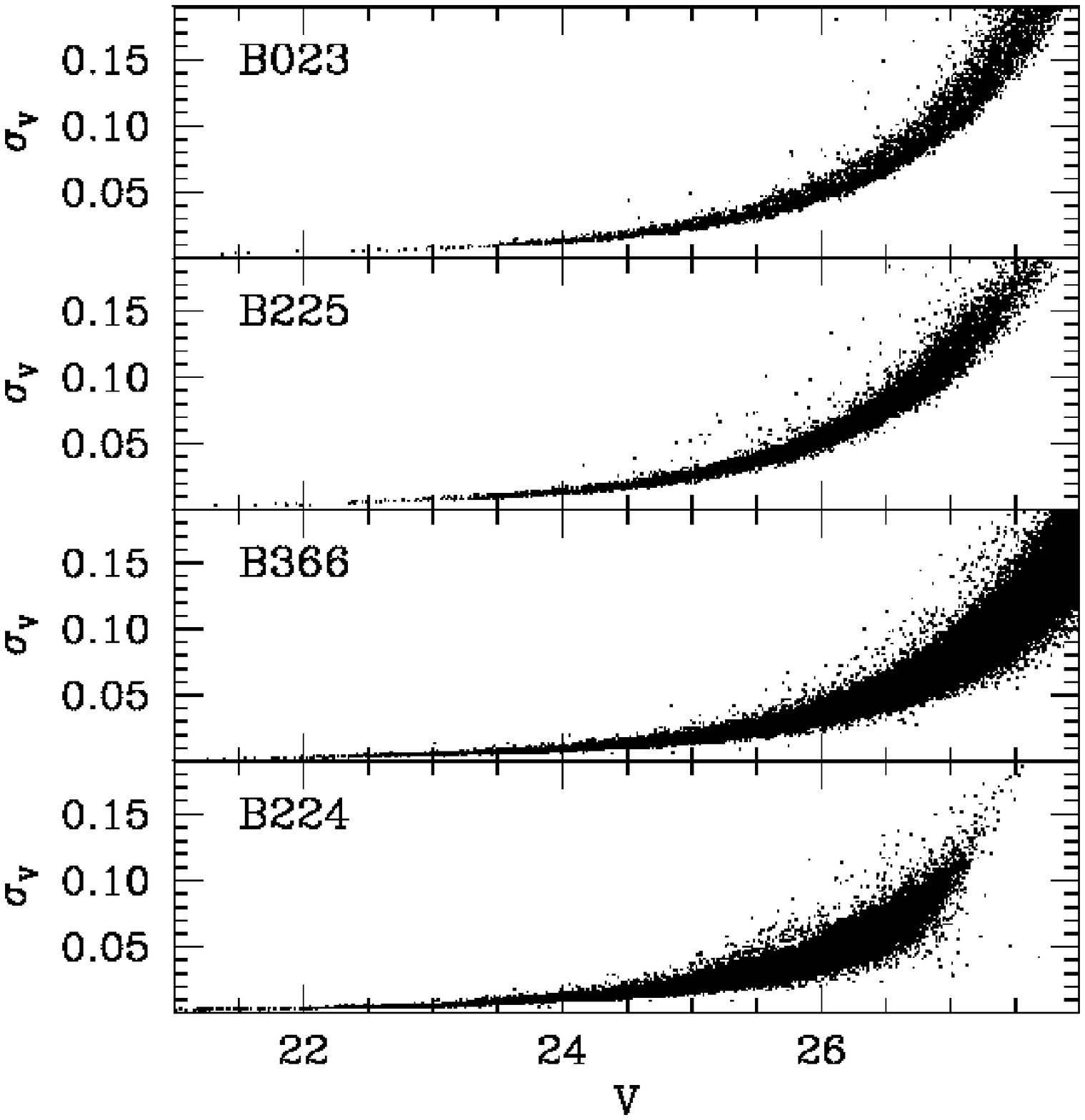}
\caption{Internal photometric errors as a function of V magnitude for 4 representative clusters,
two observed with ACS/HRC (B225 and B023), and two observed with ACS/WFC (B224 and B366). }
\label{f:errV}
\end{figure}

The  limiting  magnitudes  of our  photometry  range  from  V$\sim$26 for the
fields observed with relatively short exposure times, to  V$\sim$27.5 for the
deepest ones. The internal photometric errors of individual measures are in
general within  the range 0.01 - 0.08 mag for stars brighter than V=26,
 (see Fig. \ref{f:errV}) depending  quite strongly on the degree of crowding. 
 However, errors increase
rapidly for fainter  stars, along with the impact of blending.  Since we are
mainly interested in the position and morphology of the main CMD branches we
have not performed artificial stars experiments to study in detail  the
completeness of the samples as a function of magnitude.  However, based on
simple tests and on our previous experience, we are confident  that in all of
the considered cases the completeness is more than sufficient  ($\ga$70\%) to
achieve our scientific goals for V$\la$26.


To have an idea of the characteristic sizes of the clusters we estimated 
half-light radii -- R$_{h}$ (see Table \ref{t:parme})  by aperture photometry over concentric
annuli centered  on the cluster and extended out to sufficiently large
distances to  properly sample the background. This approach is quite  rough,
nevertheless the values obtained here for the 5 clusters (B023, B088, B158, 
B225, B407)  in common with  Barmby et al. (2007) agree within 0.05
arcseconds (i.e. to better than 0.2 pc at the M31 distance) in all cases.

\begin{figure*}
\centering\includegraphics[width=14.cm]{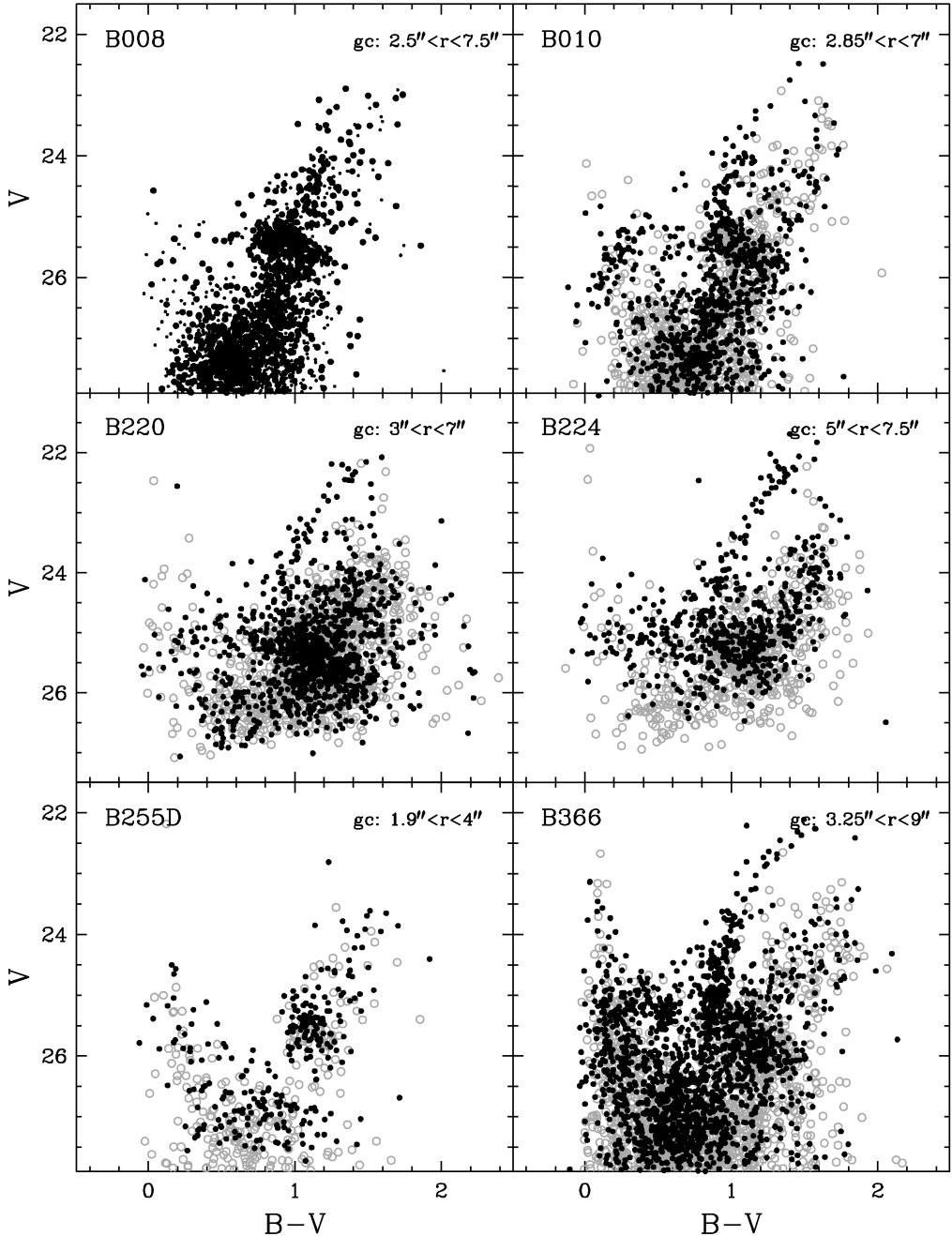} 
\caption{The CMDs of the target GCs B008, B010, B220, B224, B255D,
and  B366. {\it Filled black circles}  
are stars measured within the annulus with radii $ r$ in arcsec 
from the  cluster centre (as reported in each panel). They are taken to 
represent the cluster population; {\it open grey circles } are stars 
measured  within an outer area, of the same size, around the cluster,
and represent the surrounding field population. 
}
\label{f:cmd1}
\end{figure*}

\begin{figure*}
 \centering\includegraphics[width=14.cm]{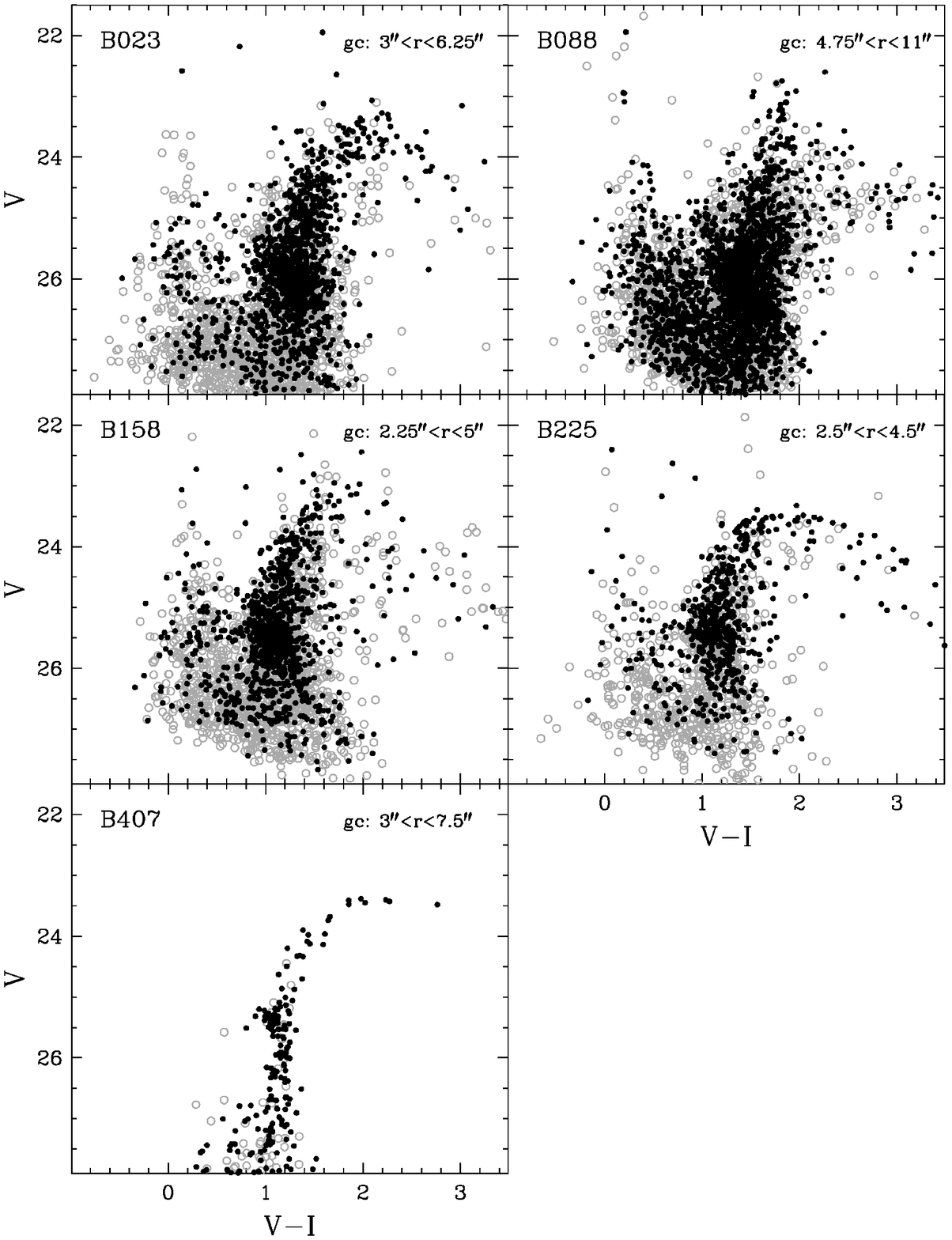} 
\caption{Same as in Fig. \ref{f:cmd1} for the GCs 
B023, B088, B158, B225, and  B407.   
}
\label{f:cmd2}
\end{figure*}

The individual  CMDs are shown in Figs. \ref{f:cmd1} and \ref{f:cmd2},  where the
cluster and field stellar populations are indicated with different  symbols
(filled black and open grey circles, respectively).  The cluster CMDs shown in
these figures sample the stellar population within an annulus around the cluster 
centre where the cluster members are more readily distinguishable with respect to the
surrounding field.  The inner limit of the annulus is set by the crowding level
that prevents from performing useful photometry in the most central region of the
cluster, the outer limit is set by the limiting radius of the cluster and by the
need to avoid contamination by the surrounding field population. The inner and 
outer radii of the adopted annuli are indicated for each cluster.  The field
population is measured on an outer concentric annulus having the  same area as
the cluster annulus. In all the CMDs shown in Figs. \ref{f:cmd1}  and \ref{f:cmd2}
the cluster population can be distinguished from the field. In most cases the
clusters show a thinner and much steeper RGB with respect to the field, and in
many cases a Blue HB is visible, that has no (or much weaker) counterpart in the 
field population. 

Before proceeding with the analysis of the cluster properties (discussed in
Sect.~3), we have applied the field decontamination procedure  described in
Bellazzini  et al. (1999). This method is based on a clipping routine which,
making use of the local density on the CMDs of the field and of the cluster,
computes the probability that a given star is a member of the cluster and retains
or rejects stars from the cluster CMD according to that.
 To verify the reliability of this procedure we carried out several decontamination tests
using different areas of the field and different techniques. In particular we applied
to the most contaminated clusters a statistical subtraction procedure based
on a Monte Carlo approach, where up to 5000 field-subtraction trials 
were used, thus obtaining globular cluster measured samples weighted by 
a statistical membership likelihood.
 Figs. \ref{f:ridge1} and
\ref{f:ridge2}  show that the decontamination of our primary targets was quite
successful, providing ``clean'' CMDs in which the main cluster branches 
are more clearly identified (the individual cases are briefly
discussed in Sect.~5). Therefore, the following analysis is based on
the decontaminated CMDs.

\subsection{Comparison with Fuentes-Carrera et al. (2008) photometry}

\begin{figure}
 \includegraphics[width=9.cm]{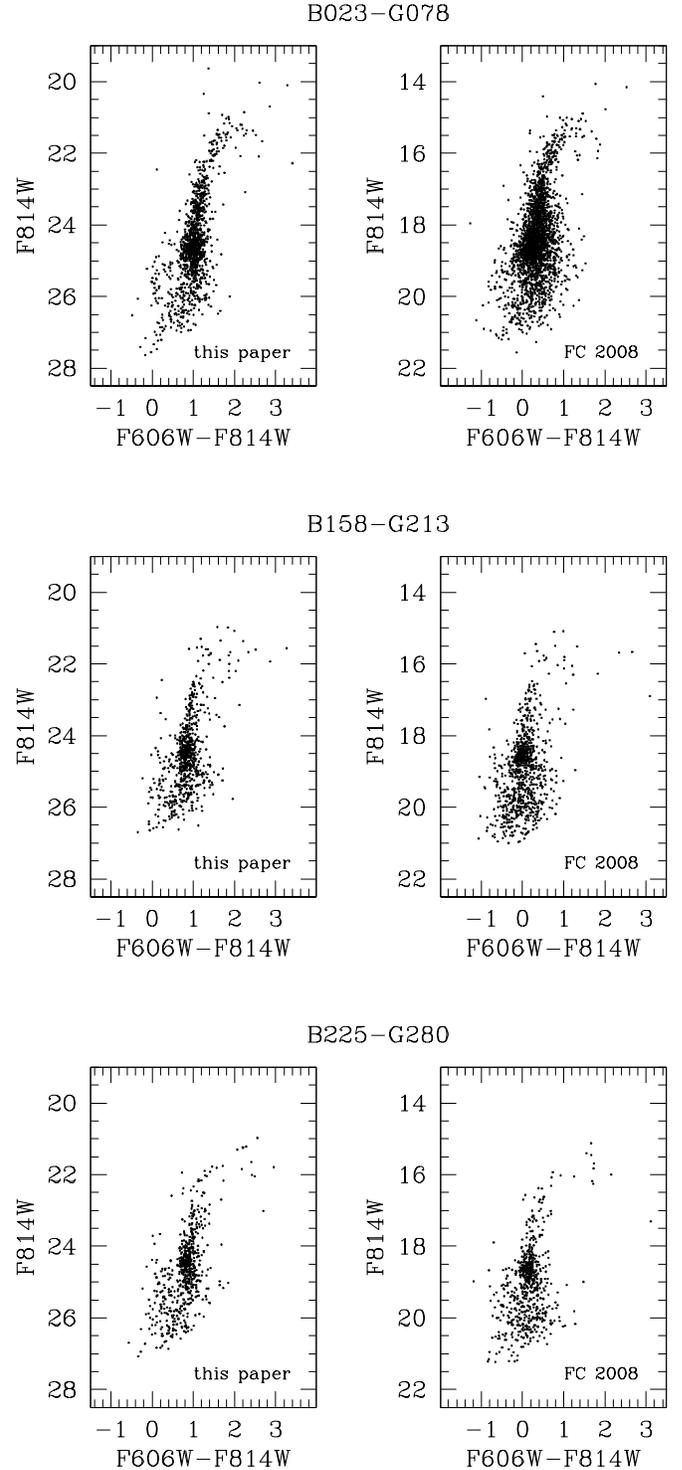}
\caption{Comparison of the CMDs obtained from the present study (left) and by Fuentes-Carrera 
et al. (2008, their Fig. 6) (right, uncalibrated VEGAMAG magnitudes), for the clusters B023 (top), 
B158 (middle) and B225 (bottom). 
 }
\label{f:conf1}
\end{figure}

While carrying out the present analysis, independent photometry of three 
objects included in our primary sample 
(B023=G078, B158=G213, B225=G280) was produced by Fuentes-Carrera  et al.
(2008) based on the same data set.  
Both CMDs for each of
these three clusters are shown side by side in Fig. \ref{f:conf1}, showing
an excellent degree of consistency in magnitude  and colour
extension and in the quality of individual star photometry.   
The close coincidence of the main
branches and even of most of the detected stars testifies the strict similarity
and agreement of these two independent photometries.  

Fuentes-Carrera et al. have focussed their analysis on the claimed 
existence of metallicity spreads in these very bright and populous GCs, based on
the  intrinsic width of the main branches. Although the quality of their data
reduction is comparable to ours,  we have not dealt with this aspect which is
beyond the scope of the present study.  We refer the interested reader to their
work for a detailed discussion of this topic.



\section{M31 vs. Galactic GCs: direct comparisons of the CMDs}  

We estimate the distance, metallicity, and reddening of
our primary clusters by comparison with a set of CMD templates of well
studied Galactic GCs, similarly to Rich et al. (2005), and Mackey
et al. (2006, 2007). Relying on the hypothesis that the considered clusters are
of similar nature as their Galactic counterparts we searched for the set of
parameters ($(m-M)_0$, E$(B-V)$ and [Fe/H]) producing the best match between the
observed RGBs and HBs and the ridge lines  of the template clusters in the
absolute plane, given the direction of the reddening vector 
$A_V=3.1$E$(B-V)$,  $A_I=1.94$E$(B-V)$  and E$(V-I)=1.375$E$(B-V)$ (Schlegel et al. 1998).

The best match was judged by eye guided by (extensive) experience, as this
approach is much more robust than most automated algorithms in presence of
significant residuals from the decontamination procedure. The steepness of the RGB
is of great help in judging if the branch is red because of high metallicity or
because of high reddening; the fact that the HB match is mostly sensitive to
vertical (magnitude) shifts, while the RGB is mostly sensitive to horizontal 
(colour) shifts also provides a useful guide to the solution. 
Colour and magnitude shifts are applied iteratively until a satisfactory 
match with any RGB and HB template is found: from these shifts  
we obtain estimates of the reddening  and distance, while  the metallicity is 
estimated by interpolation between  the two RGB ridge 
lines bracketing the observed RGB locus. 


\begin{table}
\centering
\caption{Reference grid of template Galactic globular clusters.}
\begin{tabular}{@{}lcccc@{}}
\hline
\\
 ID & [Fe/H] & E$(B-V)$ & $\mu_V$ & Phot. \\
    & dex    &  mag    &  mag    &       \\
\hline
\\
NGC7078 (M15)  & --2.16 & 0.10 & 15.37  & BV \\
NGC6397        & --1.91 & 0.18 & 12.36  & VI \\ 
NGC5824        & --1.87 & 0.13 & 17.93  & BV \\
NGC5272 (M3)   & --1.66 & 0.01 & 15.12  & VI \\ 
NGC6205 (M13)  & --1.65 & 0.02 & 14.48  & BV \\
NGC5904 (M5)   & --1.40 & 0.03 & 14.46  & VI,BV \\
NGC6723        & --1.12 & 0.05 & 14.85  & BV \\
47 Tuc         & --0.71 & 0.04 & 13.37  & VI,BV \\
NGC6624        & --0.35 & 0.28 & 15.36  & BV \\
NGC6553        & --0.29 & 0.63 & 15.83  & VI \\
\\
\hline
\end{tabular} 
{\begin{flushleft}{NOTES: Metallicities are from Zinn (1985); all other parameters are from 
Harris (1996) (online update  2003). V,I photometry is from Rosenberg et al.(2000a,b);
B,V photometry is from Piotto et al. (2002).}
\end{flushleft}}
\label{t:ggc1}
\end{table}


As starting values for the iterative procedure  we have  used E$(B-V)=0.08$ for 
the foreground reddening (Barmby et al. 2007;  Burstein and Heiles 1984), 
and the distance modulus $\mu_0=24.47$ mag for all the M31 clusters 
(McConnachie et al. 2005).
The ridge lines of the reference GGCs were assembled from the observed 
CMDs (Piotto et al. 2002 for BV photometric data, and Rosenberg et al. 2000a,b 
for VI) that were shifted to the 
absolute reference frame by correcting for reddening and distance 
using the values listed in  Table \ref{t:ggc1}. 
These reference GGCs have been chosen to provide a sufficiently fine and 
regular sampling over a wide enough range of metallicities for a correct 
characterization of the target GCs. 

In Figs. \ref{f:ridge1} and \ref{f:ridge2} we show the field decontaminated CMDs and, 
overplotted, the reference grid  of GGC ridge lines, where the bracketing RGB 
reference clusters are highlighted. 
The values of metallicity, 
reddening and distance corresponding to the best match are also reported in each
individual panel, as well as in Table \ref{t:parme};
the typical uncertainty on the distance modulus is $\pm 0.2$ mag, $\pm 0.04$ mag in
E$(B-V)$, and $\pm 0.25$ dex in metallicity.
We think that the solutions presented in Fig.~\ref{f:ridge1} and Fig.~\ref{f:ridge2}
are satisfactory and reliable. We have explored also alternative solutions,
some of which are discussed in Sect.~5.
In all cases the final adopted solution was the one which provided  
the best fit for both RGB and HB simultaneously.

As a matter of fact, due to the intrinsic and well-known age-metallicity
degeneracy, also age could be considered as an additional free parameter, which
would further complicate the analysis, having a (minor) effect on the colour of
the RGB. Since the data are not deep enough (i.e. to the main sequence
turn-off) to allow us to estimate the  cluster ages (for ages 
larger than $\sim 2$ Gyr), we have assumed that  all  of the 11 primary target
are classical old globulars (i.e. age $>10$ Gyrs). This assumption is supported
by the overall morphology of the CMDs, in particular for those clusters
displaying a Blue HB.

The best fitting procedure allowed us to estimate also the mean apparent V
magnitude of the  HB, V(HB),   by reading the value of the HB apparent
magnitude level directly on the  {\em adopted HB ridge line}  at 
(B--V)$_0$=0.3  or (V--I)$_0$=0.5  for the  metal-poor clusters. This colour has been
chosen to represent the middle of the instability strip. For the
metal-rich clusters  we have estimated V(HB) at the blue end of  the red HB
clump, with an additional correction of 0.08 mag
to recover the  mean level of the HB at the colour of the corresponding instability strip
(see Fusi Pecci et al. 1996). The
uncertainties affecting the  V(HB) estimates are often quite large, due to the 
intrinsic quality of the available data and the possible residual field 
contamination.  We have conservatively adopted $\pm$ 0.15 mag for all the
considered clusters. V(HB) and M$_V$(HB) are reported in Table \ref{t:parme},
together with the other parameters derived  from the above procedure. 

In the following section we briefly discuss the cases of each individual
cluster. 
                                                                                                   

\section{Comments on the individual clusters}  


\subsection{B008 = G060} \label{B008}

In spite of the strong field contamination the typical cluster morphology 
can be identified in the decontaminated CMD of B008. 
The cluster displays a red HB and an RGB falling about halfway  between the 
ridge lines of  47 Tuc and M5, with no need of adjustment with respect to the 
initial assumptions on distance and reddening. This leads to estimate a metallicity 
[Fe/H] $= -1.0\pm0.25$ (the error is the typical uncertainty in the
interpolation  between  the bracketing ridge lines).  
This result is in marginal disagreement (at $<2\sigma$ level) with the
estimates by Perrett et al. (2002, hereafter P02; 
[Fe/H] $= -0.41\pm 0.38$), and by Galleti et al. (2009, hereafter G09; 
[Fe/H] $= -0.47\pm 0.35$), both obtained from integrated ground-based spectroscopy. 
We collect in Table \ref{t:metal} all the available metallicity determinations 
for all the target clusters, for convenience of comparison with the present estimates. 
On the other hand,
adopting the reddening E$(B-V) = $0.21  (as estimated by Barmby et al. 2000, and private communication, 
hereafter B00)  an acceptable fit to the CMD morphology
could only be obtained for  $\mu_0 = $24.20  and [Fe/H] $ = -1.8$, with even
larger  disagreement with the spectroscopic metallicity estimates. 
Although this solution cannot be excluded in principle, we consider it as highly
unlikely, as our adopted best values provide a much better fit to the observed
CMD.


\subsection{B010 = G062}  \label{B010}
 
In this case, the decontaminated CMD is quite clean, showing a well defined and
populated  Blue HB  and a steep RGB, indicating old age and low metal content. 
The best match of these features with the corresponding ridge lines  is
obtained by assuming a value of reddening E$(B-V)= 0.18$ mag and a   
distance modulus $\mu_0 = 24.25$.
The solution relies on the best match to the blue part of the HB,
considering the handful of (supposed) HB stars around $0.3 \la $(B--V)$_0\la 0.5$ as
{\em evolved} BHBs, i.e. post--ZAHB stars in their way to the 
Asymptotic Giant Branch (and hence brighter than the genuine unevolved HB stars that
we are using as standard candles).  

\begin{figure*}
 \centering\includegraphics[width=13.5cm]{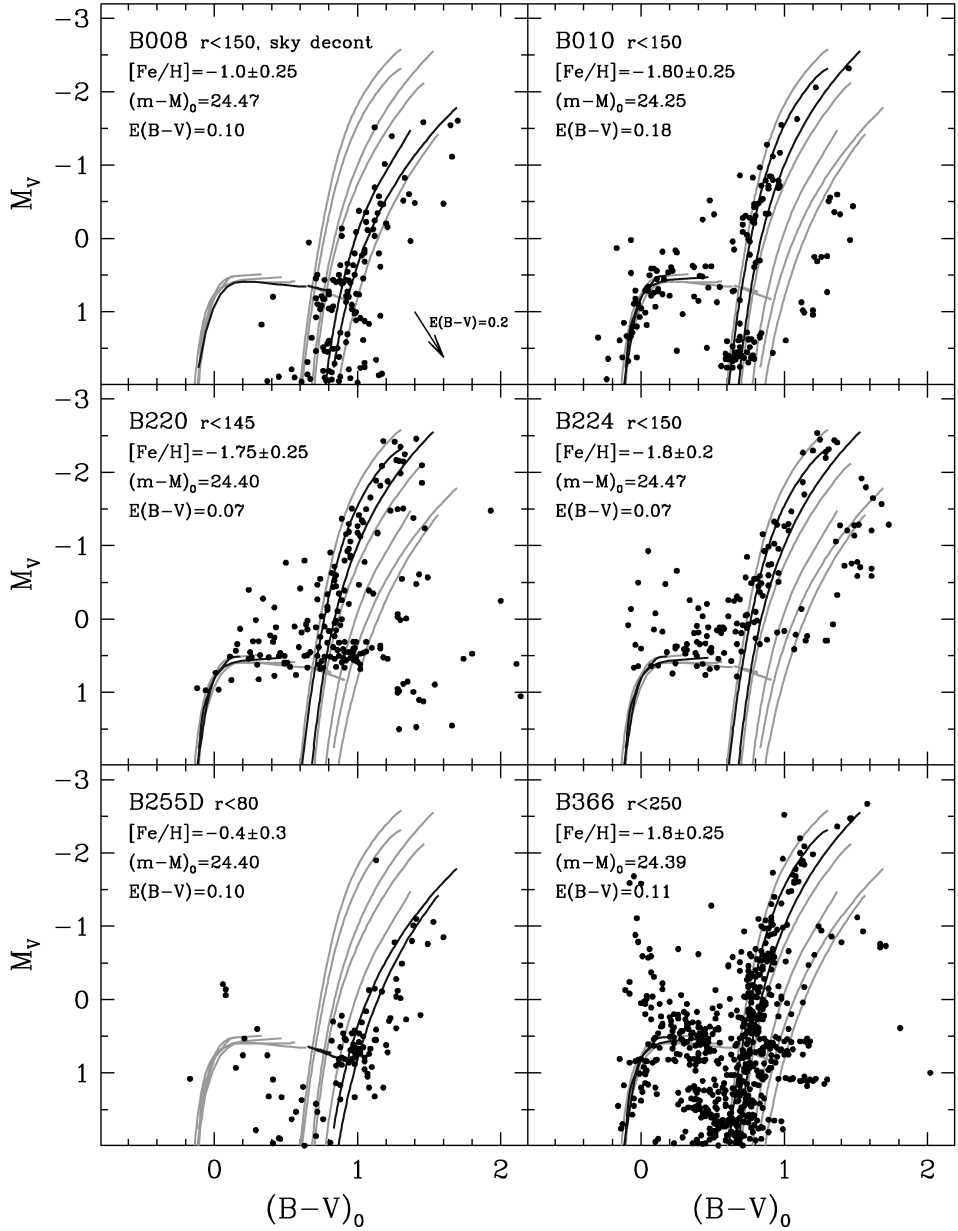} 
\caption{The CMDs of the primary target GCs B008, B010, B220,
B224,  B255D and B366. The data have been decontaminated by the field
contribution. The arrow in the left top panel indicates the reddening vector
corresponding to e$(B-V)=0.2$.
The  {\it bracketing ridge lines} of reference Galactic GCs are
also shown,  as described in Sect. 4 and Table \ref{t:ggc1}.
}
\label{f:ridge1}
\end{figure*}

\begin{figure*}
 \centering\includegraphics[width=13.5cm]{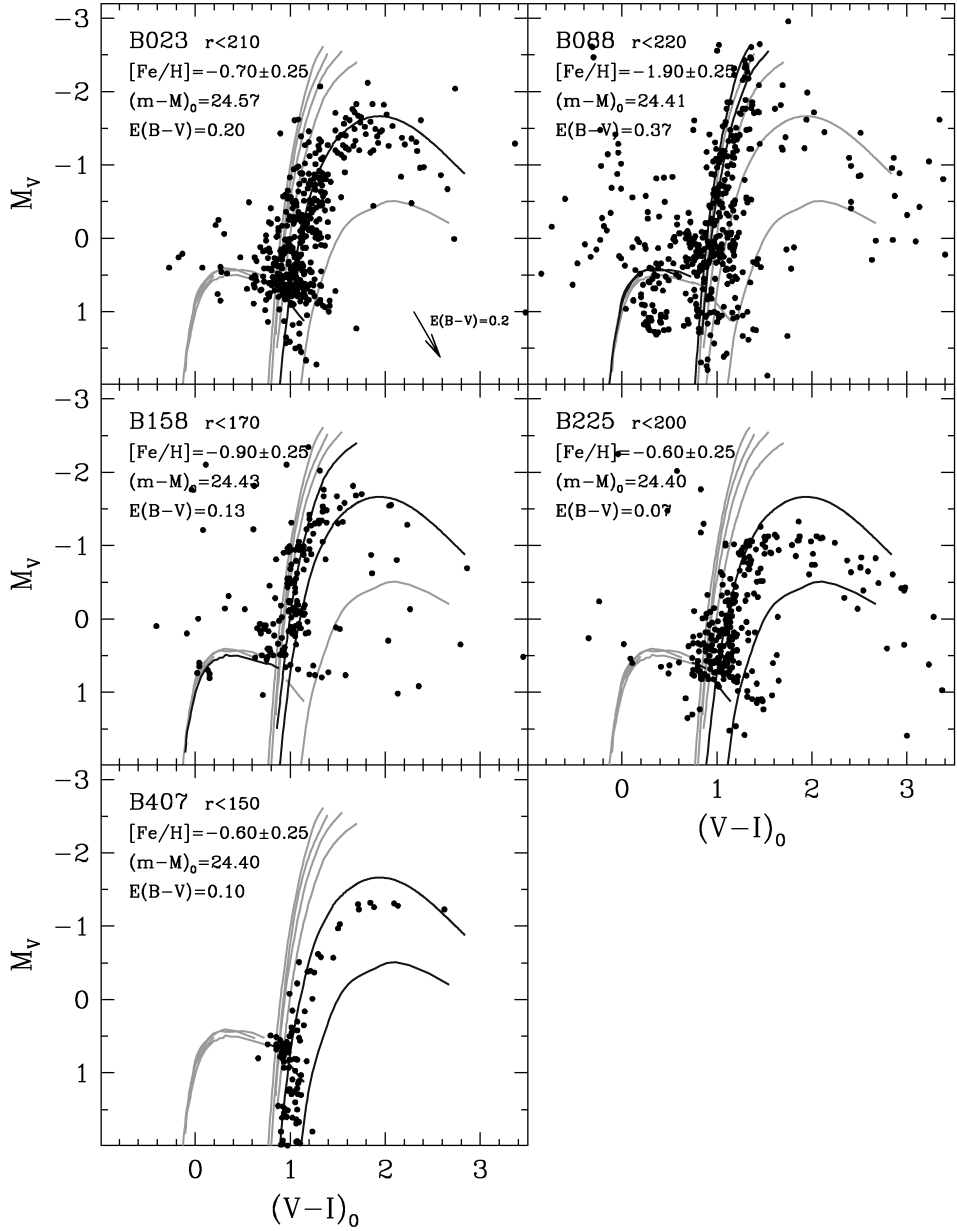} 
\caption{Same as in Fig. \ref{f:ridge1} for the GCs 
B023, B088, B158, B225 and B407.   
}
\label{f:ridge2}
\end{figure*}

With these assumptions, the CMD shown in Fig. \ref{f:ridge1} indicates that
the  metallicity of B010 is very similar to NGC5824, namely 
[Fe/H] $= 1.8\pm0.25$.  This value is in good agreement with  the spectroscopic
ground-based estimates, [Fe/H] $ = -1.87\pm 0.61$ (Huchra et
al. 1991, hereafter HBK), [Fe/H] $ = -1.77\pm 0.14$ (P02), and 
[Fe/H] $ = -1.64\pm 0.68$ (G09).    Also the adopted
reddening E$(B-V)=0.18$ is fully consistent with the values reported in the
literature, i.e. 0.19 (B00) and 0.22 (Fan et al. 2008, hereafter F08). 


\subsection{B220 = G275}  \label{B220 = G275}

The CMD of B220 shows the presence of a well defined BHB and a rather steep
RGB, indicating old age and a low metallicity  content.  The best match of
these features of the CMD with the corresponding reference  ridge lines is
obtained by assuming a value of reddening E$(B-V)=0.07$  mag (in agreement with
E$(B-V)=0.05$ by F08, and E$(B-V)=0.08$ by B00) and a distance modulus $\mu_0=24.40$.
 With these assumptions 
the CMD  shown in Fig. \ref{f:ridge1} indicates that the metallicity of B220 is
intermediate  between M13 and NGC5824,  [Fe/H] $=-1.75\pm0.25$. 
This value compares fairly well with the spectroscopic  estimate of HBK,
[Fe/H] $=-2.07\pm 0.82$, whereas the values found by P02,  [Fe/H] $=-1.21\pm
0.09$  and G09 [Fe/H] $=-1.09\pm 0.42$ seem too high for this cluster. 


\subsection{B224 = G279}    \label{B224}

The best match of the steep RGB and extended HB of B224 with the corresponding
reference ridge lines is  obtained by assuming a value of reddening
E$(B-V)=0.07$ mag and the standard  distance modulus of 24.47 mag.   With
these values, the CMD shown in Fig. \ref{f:ridge1} indicates that the 
metallicity of B224 is intermediate between M13 and NGC5824, 
[Fe/H] $=-1.80\pm0.25$. This value compares well with previous
estimates from integrated spectroscopy: [Fe/H] $=-1.90\pm0.24$ (HBK),  
[Fe/H] $=-1.80\pm0.05$ (P02),  and
[Fe/H] $=-1.68\pm0.28$ (G09). 

Both F08 and  B00 have estimated slightly higher reddening values: 0.13 and
0.12 mag,  respectively. We have searched for solutions
with E$(B-V)=0.13$, and we found that the best fit would yield a similar
metallicity but a much  shorter distance,  $\mu_0=24.25$. However, the
overall quality of the fit is significantly  worse when using this higher
value of reddening, so we have adopted our  primary solution. 


\subsection{B255D}  \label{B255D}

The cluster is
rather  small and the statistical decontamination procedure becomes less
effective when  the number of stars is low.  As a result, one
can still see the presence of some residual field population on  the blue side
of the CMD (blue plume).  Nevertheless, a sparse
and metal-rich RGB as well as a red clump can be seen
clearly.  
The best match with the ridge lines in this case is not much more than an
intelligent  guess, and indicates a metallicity [Fe/H] $=-0.40\pm0.25$ and a
distance  modulus  $\mu_0=24.40$ mag  for the  assumed value
of reddening E$(B-V)=0.10$ mag. 
There are no ground-based spectroscopic estimates for this cluster. 


\subsection{B366 = G291}  \label{B366}

B366 is a rather populous cluster lying in a high density 
field, as shown in Fig. \ref{f:gc}. The decontamination procedure was not
able to eliminate completely the field component (a blue plume as well as 
a red clump to the red of the cluster RGB), but the cluster
population shows up quite clearly as a
well defined  HB with a possible blue extension, and a rather steep RGB,
suggesting old age and metal deficiency.
The cluster is classified as old also by  C09, based on
its integrated spectrum.  

The best match between the observed CMD and the template ridge lines
is  achieved with E$(B-V)=0.11$ mag and  $\mu_0=24.39$ mag.  
With these values, [Fe/H] $=-1.80\pm 0.25$  is found.  
This value is consistent, within the uncertainties, with the 
spectroscopic estimates by HBK, [Fe/H] $=-1.39\pm 0.28$, and
G09, [Fe/H] $=-2.14\pm 0.39$, while it is in excellent agreement with 
the results of P02,  [Fe/H] $=-1.79\pm 0.05$. 


\subsection{B023 = G078}   \label{B023}

The field decontamination has left some marginal field contribution on  the
bluest part of the CMD,  but the main branches  stand out
quite clearly. The cluster has a red HB, and its RGB
falls almost exactly on the ridge line of 47~Tuc.

The best match of the main branches is obtained for 
E$(B-V)=0.20$ mag and $\mu_0 = 24.57$. This leads to estimate a metallicity 
[Fe/H] $=-0.70\pm0.25$, in good agreement with existing spectroscopic estimates,
[Fe/H] $=-0.92\pm0.10$ by HBK, and [Fe/H] $=-0.91\pm0.15$ by G09.  

We note that the reddening estimated by Barmby et al. (2007, hereafter B07),
and F08 is significantly larger,  E$(B-V)=$0.36 and 0.32  mag, respectively.
With these values   no match can be achieved with any of the ridge lines,
therefore we exclude the  possibility of such a high reddening for this
cluster.


\subsection{B088 = G150}  \label{B088}

As one can see in Fig. \ref{f:gc}, this cluster is very populous, has a 
strongly elliptical shape and lies in a rather dense field.  Two other clusters
in our sample, B023 and B366, show some evidence of elliptical  shape, but the
ellipticity of B088 is clearly larger. The values reported in the literature
are $\epsilon=0.28$ (Barmby et al. 2007),  $\epsilon=0.18$ (Staneva et al.
1996) and $\epsilon=0.23-0.27$ (Lupton 1989),  making this object particularly
noteworthy.

In this case, where the stellar field is very crowded and variable, we have
performed several statistical field subtraction experiments. 
In spite of the presence of some residual contamination from the field, the
steep cluster RGB is clearly identified in all cases, indicating a low metal 
content. On the other hand the HB
morphology is more confused, and the vertical match is rather tentative. A
possible adopted set of parameters is  [Fe/H] $=-1.90\pm0.25$, E$(B-V)=0.37$  and
$\mu_0=24.41$. The metallicity agrees very well with spectroscopic estimates,
[Fe/H] $=-2.17\pm0.48$ (HBK),  [Fe/H] $=-1.81\pm0.06$ (P02), and
[Fe/H] $=-1.94\pm0.52$ (G09).  
A high value of reddening for this cluster was found independently by  F08
(0.46 mag) and B07 (0.48 mag). Our result indicates that the cluster is 
located in the nearest side of the M31 disc, and lies behind some dust
layer as clearly visible in the Spitzer images of this region (Gordon et al.
2006).


\subsection{B158 = G213}  \label{B158}

Even if sparsely populated, the steep RGB of B158  stands out quite clearly 
in the decontaminated CMD, while the fit to a (supposed) extended HB is just
tentative. Our best solution gives an estimate of the reddening
E$(B-V)=0.13$ mag, in excellent agreement with the results by F08, E$(B-V)=0.14$,
and B00, E$(B-V)=0.12$. The adopted distance modulus is $\mu_0=24.43$, and the 
metallicity [Fe/H] $=-0.90\pm0.25$, which
compares very well  with all the ground-based estimates: 
[Fe/H] $=-1.08\pm0.05$ (HBK), [Fe/H] $=-1.02\pm0.02$ (P02), and  
[Fe/H] $=-0.74\pm0.15$ (G09).


\subsection{B225 = G280}  \label{B225}

The RGB and red HB of the cluster 
stand out very clearly and are well consistent with the ridge lines of the 
metal-richest templates, on the assumption of a  reddening value 
E$(B-V)=0.07$ and a distance $\mu_0=24.40$.
This leads to estimate a metallicity 
[Fe/H] $=-0.60\pm0.25$, in agreement with the spectroscopic estimates:
[Fe/H] $=-0.70\pm0.12$  (HBK), [Fe/H] $=-0.67\pm0.12$ (P02), 
and [Fe/H] $=-0.35\pm0.15$ (G09).

The CMD of this cluster was previously obtained by  Fusi Pecci et al. (1996),
with HST/FOC and, subsequently, by Rich et al. (2005), with HST/WFPC2. Both
studies obtained results in good agreement with those presented here.


\subsection{B407 = G352}   \label{B407}

The cluster B407 lies at a rather large projected distance from the centre of 
M31, in a low density region where the contamination by field stars is very
low. As a consequence, the RGB and red HB of the cluster are very well defined.
Their position in the CMD indicates a metallicity slightly higher  
than the reference cluster 47~Tuc.
   
The best solution is obtained for E$(B-V)=0.10$ mag and  $\mu_0=24.40$ mag. 
With these values, the metallicity of B407 is [Fe/H] $=-0.60\pm0.25$, 
fully consistent with the spectroscopic estimates by HBK, 
[Fe/H] $=-0.85\pm0.33$ and, in particular,  G09, [Fe/H] $=-0.65\pm0.15$.

The case of B407 as a metal rich cluster in the outer halo of M31 
is discussed in more detail in Section 6. 

 
\begin{table}
 \centering
  \caption{Parameters derived for the 11 primary target clusters from the
 procedure described in  Sect.s 4 and 5.}
  \begin{tabular}{@{}lcc@{\hspace{0.08in}}c@{\hspace{0.05in}}c@{\hspace{0.09in}}c@{\hspace{0.09in}}c@{}}
  \hline
\\
 ID & R$_{hl}$ & V$_{HB}$&E(B-V) &$\mu_0$& [Fe/H]  &M$_V^{HB}$\\ 
    & arcs     &         &         &          & dex     &mag    \\ 
 \hline
\\
B008-G60   & 0.95 & 25.29 & 0.10 & 24.47 & --1.00$\pm$0.25 & 0.51 \\ 
B010-G62   & 1.40 & 25.30 & 0.18 & 24.25 & --1.80$\pm$0.25 & 0.49 \\ 
B023-G78   & 0.95 & 25.91 & 0.20 & 24.57 & --0.70$\pm$0.25 & 0.72 \\ 
B088-G150  & 1.11 & 25.99 & 0.37 & 24.41 & --1.90$\pm$0.25 & 0.43 \\ 
B158-G213  & 0.65 & 25.44 & 0.13 & 24.43 & --0.90$\pm$0.25 & 0.61 \\ 
B220-G275  & 2.15 & 25.08 & 0.07 & 24.40 & --1.75$\pm$0.25 & 0.46 \\ 
B224-G279  & 1.35 & 25.22 & 0.07 & 24.47 & --1.80$\pm$0.25 & 0.53 \\ 
B225-G280  & 0.61 & 25.35 & 0.07 & 24.40 & --0.60$\pm$0.25 & 0.73 \\ 
B255D-D072 & 1.60 & 25.53 & 0.10 & 24.40 & --0.40$\pm$0.25 & 0.82 \\ 
B366-G291  & 2.00 & 25.25 & 0.11 & 24.39 & --1.80$\pm$0.25 & 0.52 \\ 
B407-G352  & 0.80 & 25.41 & 0.10 & 24.40 & --0.60$\pm$0.25 & 0.70 \\ 
\\
 \hline                                                                                                  
\end{tabular}                                                                                                  
\label{t:parme}
\end{table}


\begin{table}
 \centering
 \caption{Comparison of the estimates of metallicity here obtained for the target clusters
(see Sect.s 4 and 5) and previous recent determinations. }
 \begin{tabular}{@{}l@{\hspace{0.09in}}c@{\hspace{0.08in}}c@{\hspace{0.08in}}c@{\hspace{0.08in}}c@{}}
 \hline
\\
 ID	      & [Fe/H]$_{CMD}$ & [Fe/H]$_{G09}$ &[Fe/H]$_{P02}$ &[Fe/H]$_{HBK}$\\ 
	      &  dex           &  dex 	          & dex             &dex             \\ 
 \hline
\\
  B008-G60    & --1.00$\pm$0.25 &  --0.47$\pm$0.35 & --0.41$\pm$0.38  & 		  \\				     
  B010-G62    & --1.80$\pm$0.25 &  --1.64$\pm$0.68 & --1.77$\pm$0.14  & --1.87$\pm$ 0.61 \\
  B023-G78    & --0.70$\pm$0.25 &  --0.91$\pm$0.15 &		      & --0.92$\pm$ 0.10 \\ 
  B088-G150   & --1.90$\pm$0.25 &  --1.94$\pm$0.52 & --1.81$\pm$0.06  & --2.17$\pm$ 0.48 \\ 
  B158-G213   & --0.90$\pm$0.25 &  --0.74$\pm$0.15 & --1.02$\pm$0.02  & --1.08$\pm$ 0.05 \\ 
  B220-G275   & --1.75$\pm$0.25 &  --1.09$\pm$0.42 & --1.21$\pm$0.09  & --2.07$\pm$ 0.82 \\ 
  B224-G279   & --1.80$\pm$0.25 &  --1.68$\pm$0.28 & --1.80$\pm$0.05  & --1.90$\pm$ 0.24 \\ 
  B225-G280   & --0.60$\pm$0.25 &  --0.35$\pm$0.15 & --0.67$\pm$0.12  & --0.70$\pm$ 0.12 \\ 
  B255D-D072  & --0.40$\pm$0.25 &		  &		      & 		 \\
  B366-G291   & --1.80$\pm$0.25 &  --2.14$\pm$0.39 & --1.79$\pm$0.05  & --1.39$\pm$ 0.28  \\
  B407-G352   & --0.60$\pm$0.25 &  --0.65$\pm$0.15 &		      & --0.85$\pm$ 0.33  \\
\\
 \hline 												 
\end{tabular} 
{\begin{flushleft}{CMD: this paper; G09: (Galleti et al. 2009); P02: Perrett et al. (2002);
HBK: Huchra et al. (1991).}
\end{flushleft}}
\label{t:metal}
\end{table}


\begin{figure*}
  \centering
  \includegraphics[width=13.5cm]{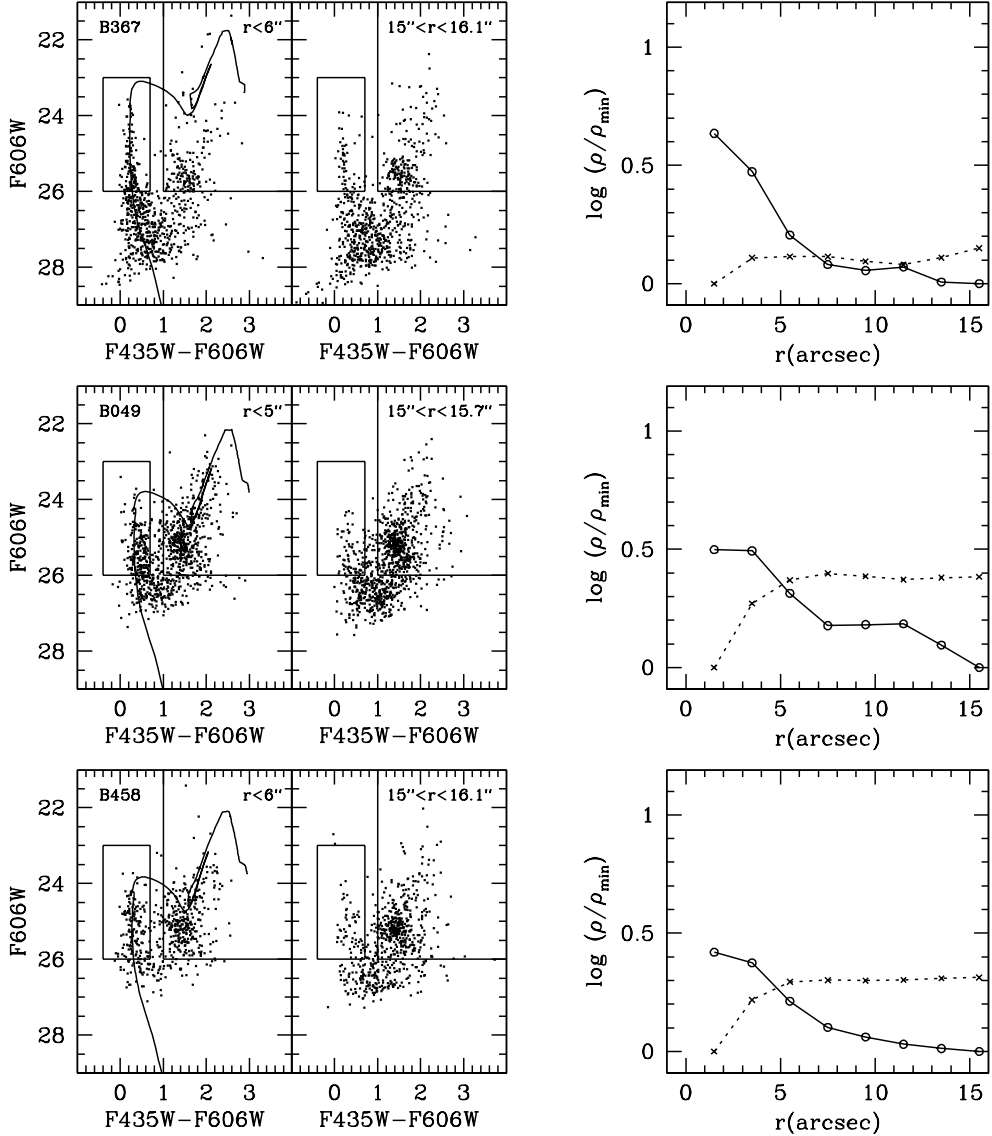}
\caption{$left$: CMDs of annuli dominated by the cluster population; $right$:
CMDs of the surrounding fields, measured over annuli of similar area. Right
panels report the radial profiles obtained by counting stars in the two boxes
reported in the CMD plots.  The $solid$ lines show the radial behaviour of the
"blue plume" in the smaller box, presumably including most of the  cluster MS,
while the $dotted$ lines show the corresponding trend as obtained from   the
bigger boxes, presumably dominated by the field. Best-fitting 
isochrones  with solar metallicity (Girardi et al. 2002) are overimposed. 
}
\label{f:blcc1}
\end{figure*}


\begin{figure*}
  \centering
  \includegraphics[width=13.5cm]{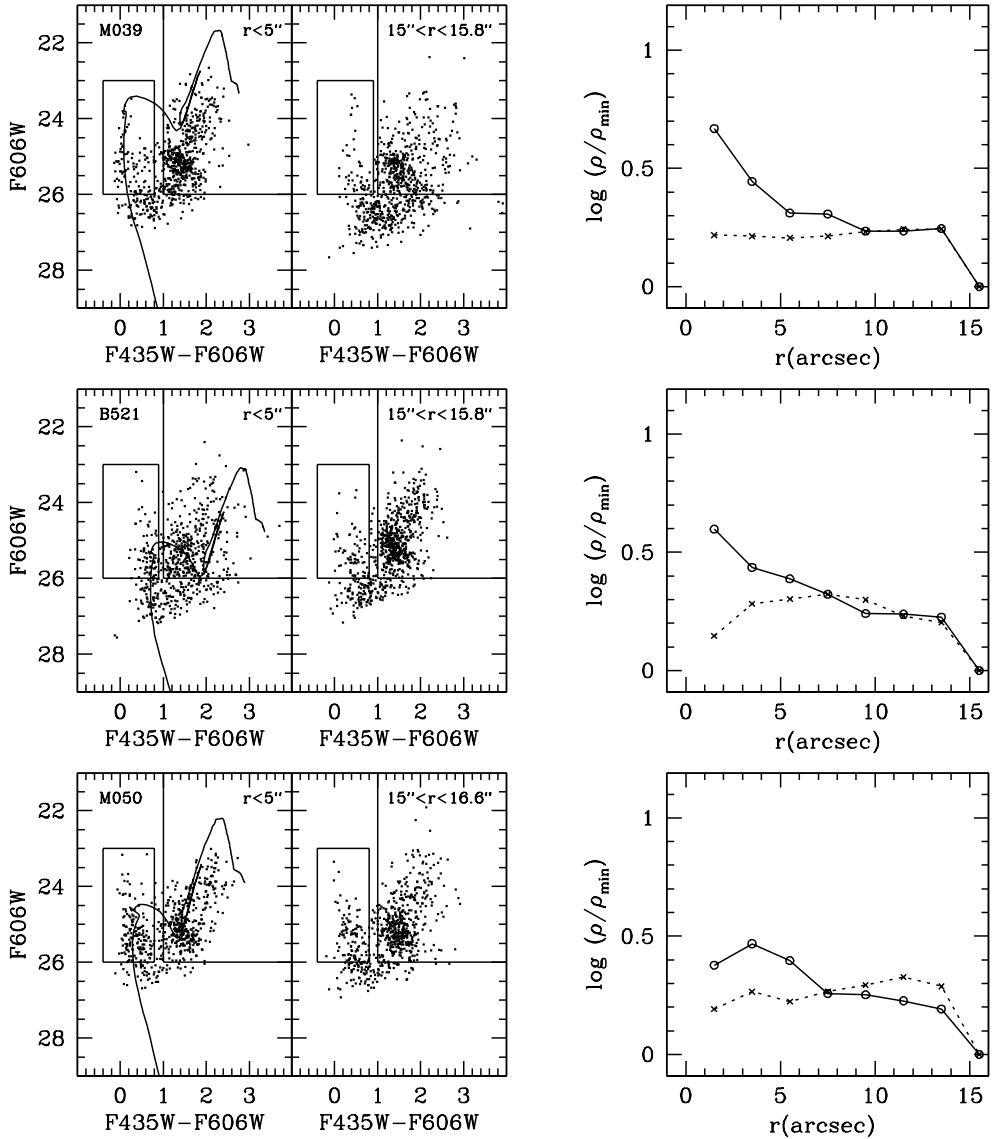}
\caption{Same as Fig. \ref{f:blcc1} for the clusters 
B521, M050 and  M039. 
}
\label{f:blcc2}
\end{figure*}


\begin{table}
  \begin{center}
  \caption{Parameters derived for the candidate young clusters. Photometry is from G04 
  except when otherwise stated. R$_{hl}$ indicates the  half-light radius. }  
  \label{t:young}
  \begin{tabular}{@{}l@{\hspace{0.05in}}c@{\hspace{0.05in}}l@{\hspace{0.05in}}c@{\hspace{0.05in}}c@{\hspace{0.01in}}c@{\hspace{0.05in}}c@{\hspace{0.01in}}c@{}}
    \hline
\\
 ID & R$_{hl}$ & ~~~V & (B-V) &E$(B-V)$&Age&E$(B-V)$&Age\\ 
    & (arcs) &      &       &             & (Myr)    &              & (Myr)    \\ 
    &          &      &       &\multicolumn{2}{c}{(~this paper~)} &\multicolumn{2}{c}{(~~ C09 ~~)}   \\ 
    \hline
\\
B049-G112 &1.20 & 17.56   & 0.52 & 0.30 & 280  & 0.25 & 400 \\
B090      &0.47 & 18.80   &      &      &      &      & old  \\
B367-G292 &0.94 & 18.45   & 0.32 & 0.25 & 200  & 0.25 & 200 \\
B458-D049 &1.60 & 17.84   & 0.49 & 0.25 & 320  & 0.25 & 500 \\
B515      &1.25 & 18.60$^1$&     &      &      &      &     \\  
B521-SK034A&0.75& 19.08$^1$&     & 0.55 & 400  & 0.38 & 250 \\  
M039      &0.62 & 18.94   &      & 0.10 & 320  & 0.18 & 320 \\  
M050      &0.80 & 18.71   &      & 0.15 & 560  & 0.25 & 300 \\  
B057-G118 &0.70 & 17.64   & 0.69 &      &      &      & old \\
\\
    \hline
\end{tabular} \end{center}
{\begin{flushleft}{
($^1$): V magnitude from C09.}
\end{flushleft}}
\end{table}


\subsection{The candidate young clusters}   \label{BLCC}

As noted in Sect.2, there are 9 clusters that we consider separately as they
have been classified as young by previous studies. Five of them, namely B049,
B057, B090, B367, B458, were included in the list  of the so-called  "Blue
Luminous Compact Clusters"   (BLCC, Fusi Pecci et al. 2005, F05 hereafter). 
They are quite
faint, V$\sim17.5 - 18.5$, but are undoubtedly clusters and some of them have
the compact appearance that is typical of GCs (see Fig.~\ref{f:blcc}; F05, 
Williams \& Hodge 2001). B057 was included by F05 among the
candidate "young" clusters  due to the quite high H$_\beta$-value, 5.56, but
C09 (see Table \ref{t:young}) classify it as "old" as well as  B090,
with a lower H$_\beta$-value, 3.38, that was included in the list of possible
young candidates by Jiang  et al. (2003).

Three other objects, B521, M050, M039 have been classified as "young" clusters
by C09 (see Table \ref{t:young}).   B521 is actually coincident
with another object, SK034A, having measured radial velocity   (v$_r= -531.8$
kms$^{-1}$, Kim et al. 2007; v$_r=-515.8$ kms$^{-1}$,  C09). M050 is classified as a 
"young" cluster by C09 who found  v$_r=-156.6$ kms$^{-1}$. It  looks like a small 
asymmetric aggregate of stars, but its CMD confirms that it is indeed a young cluster 
(see below).  M039=KHM31-516 (Krienke and Hodge 2008)  is faint and partially resolved,  
C09 list v$_r=-82.4$ kms$^{-1}$.   B515=KHM31-409 was listed by Krienke and Hodge (2008) 
as an open  cluster\footnote{We note that another cluster of our sample, B041 (not
considered in this section), that was classified as old by C09, was instead suggested 
to be young by Barmby et al. (2007).
According to the latter study  its red integrated colour is probably due to a
red, bright, non-member star  which masks the true intrinsic colour of the
cluster. Unfortunately our data do not provide any further insight on the age
of this cluster.}.


\begin{figure*}
\centering \includegraphics[width=12.cm]{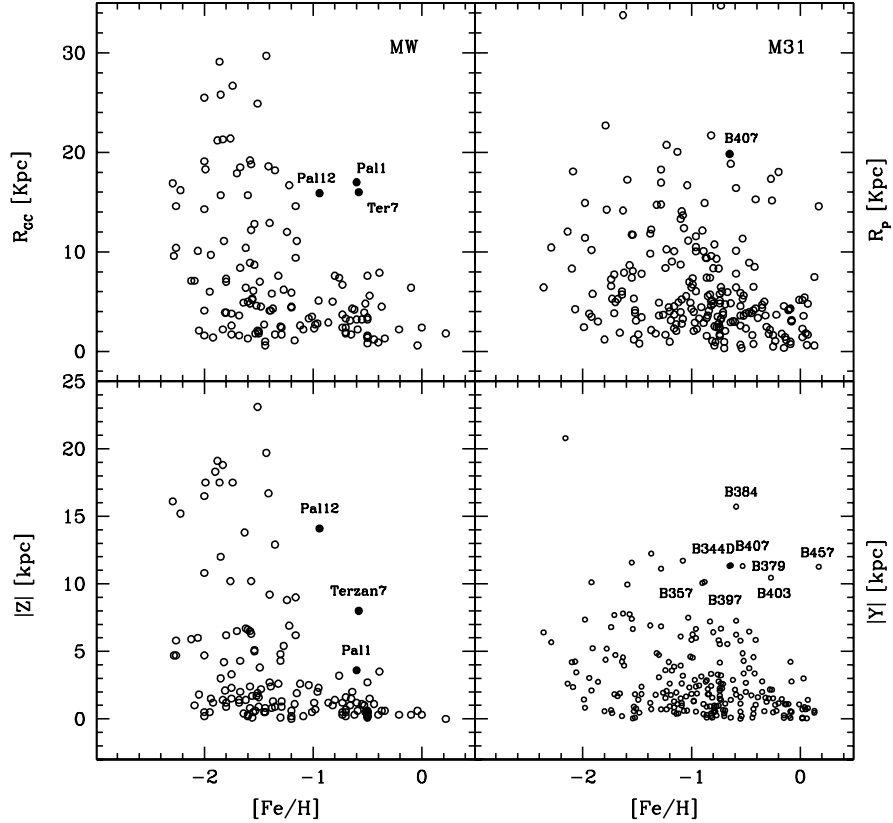} 
\caption{Left
panels: distribution of Galactocentric distance (upper-left) and absolute height
above the Galactic plane (lower-left) as a function of metallicity for Galactic
GCs (from Harris 1996). The clusters having [Fe/H]$\ge -1.0$ and R$_{GC}>10$ kpc
are plotted as filled circles and labelled with their names.   Right panels:
projected M31 galactocentric distance (upper-right) and projected distance  from
the major-axis (lower-right) as a function of metallicity for M31 CGs (from
G09). B407, as well as other clusters having [Fe/H]$\ge -1.0$ and unusually high
Y, are labelled with their names, B407 is highlighted as a filled circle. }
\label{f:rzfe} 
\end{figure*}


For 6 of the 9 clusters quoted above (B367, B049, B458, B521=SK034A, 
M039 and M050) we were able to obtain  CMDs representative of the cluster 
populations, that are shown in Figs. \ref{f:blcc1} and \ref{f:blcc2}. 
On the rightmost panels of these figures we report the cluster density
profiles obtained by counting stars on CMD boxes selecting the young main sequence 
(MS) population (open circles) and the red evolved population (RGB and Red Clump; 
crosses). 
Even if in most cases the CMD of the cluster is quite similar to that of the 
surrounding field (sampling the star-forming thin disc of M31), the density profiles 
show that in all cases a significant overdensity of MS stars is found at the 
cluster position. 
Guided by the density profiles we selected the radial annuli where the CMD is
expected to be dominated by cluster stars (leftmost panels), to be compared
with an external annulus of the same area sampling the surrounding field
(central panels).

To have a rough estimate of the age and reddening, the CMDs of the clusters
were fitted (by eye) with solar abundance isochrones (from Girardi
et al. 2002), as done in Williams \& Hodge (2001) and Perina et al. (2009a).  
The results, reported in Table \ref{t:young}, are in good agreement with similar
estimates by C09  who adopted however super-solar abundance isochrones. 
All the six clusters for which the CMD could be derived (see Figs. \ref{f:blcc1} and 
\ref{f:blcc2}) appear indeed younger than 1 Gyr, thus confirming their previous 
classification.

For the remaining three clusters B057, B090 and B515, it resulted impossible to
single out the cluster population from the background, thus we
cannot provide any improved age estimate. 

It is worth noticing, that four of the clusters considered here (B367, B049,
B458, and B057) have a spectroscopic metallicity estimate from P02
(see Tab.~2, above) that was obtained using a calibration that is based
on (and valid only for) old clusters.
Their high degree of metal deficiency reported
by P02 $-1.18<[Fe/H]<-2.32$ is very likely spurious,
due to the known
fact that a young age mimics the lack of metals in integrated colours and
spectra (see Fusi Pecci et al. (2005) for a detailed discussion of this effect
in the context of the study of the GC system of M31). Moreover, in a search for
groups of M31 GCs having common origin (from the disruption of the same parent
dwarf galaxy, for instance) based on the similarity in position, velocity, and
metallicity, Perrett et al. (2003) identified eleven remarkable groups. Their
group 9 contains B049 and B458, confirmed here as having age $<1$ Gyr from their
CMD, B057 and DAO408, classified as young from their $H_{\beta}$ and/or colour in
the RBC, and B034. Thus, four of the five members of the group are young clusters
having velocities in full agreement with
the overall rotation pattern of M31 disc. As they likely belong to the disc,
their proximity in space naturally implies similar velocities, while the
similarity in metallicity is due to their young age being mis-interpreted as low
metal content, as described above. We conclude that this proposed group does not
trace a real overdensity in the phase-space of the M31 halo, but simply
a bunch of bright young disc clusters lying in the same spot of the disc.

A thorough discussion of "young" and bright clusters in M31 with HST-based
CMDs, based on a wide homogeneous sample of other 18 candidates (P.I. J. Cohen
GO 10818) and also including the six clusters studied here and the four
clusters by  Williams and Hodge (2001), will be presented in a forthcoming paper 
(see Perina et al. 2009a, for a presentation of the overall project). 
For a discussion about faint young clusters in M31 we refer the reader 
to  Krienke and Hodge (2007, 2008).

\section{Clusters in Streams}

Among all the clusters of our sample, B407 is the most distant from the
centre of M31,  lying at a projected distance of about 20 kpc. It is also one of
the most metal rich,  and this combination makes it worth a more detailed
investigation. 

In Fig. \ref{f:rzfe} we show the distribution of Galactocentric
distance  and absolute height above the Galactic plane  as a function of
metallicity for GCs  in the Milky Way (from Harris 1996). It is quite clear
that, while metal-poor clusters ([Fe/H]$\la -1$) are found at any $R_{GC}$ and/or 
$\vert Z\vert$ , the metal-rich ([Fe/H]$\ga -1$) clusters
are confined within $R_{GC}<8$ kpc and $\vert Z\vert< 3$ kpc\footnote{Incidentally, 
we note that the transition between the clusters confined to low R$_{GC}$
and $\vert Z\vert$ and those distributed over the whole range spanned by these
parameters seems to be {\em very sharp}, occurring nearly exactly at [Fe/H]$=-1.2$.}. The only
exceptions are three
metal-rich clusters that do not satisfy these conditions and stand out as obvious
outliers in Fig.~\ref{f:rzfe}, namely Terzan~7, Palomar~12 and Palomar~1.
Ter~7 is a member of the Sagittarius
dwarf spheroidal galaxy (Ibata et al. 1994, 1995), a satellite of the
MW that is currently disrupting under the strain of the Galactic tidal field. In
this process it has developed two huge tidal tails (Sgr Stream) 
containing its former stars
(Ibata et al. 2001a; Majewski et al. 2003; Belokurov et al. 2006)
and clusters (Bellazzini et al. 2003a) escaped during various perigalactic
passages. Pal~12 is indeed associated with the Sgr Stream (Dinescu et al. 2000; 
Martinez-Delgado et al. 2002; Bellazzini et al. 2003a,b; Cohen 2004).
An extra-galactic origin has been invoked also for Pal~1, to explain its
anomalously young age (Rosenberg et al. 1998) and its unusual abundance pattern
(Venn et al. 2007; Correnti, Saviane \& Monaco, private
communication). These characteristics are shared also by Ter~7 (Buonanno et al.
1995;  Tautvaisien\'e et al. 2004; Sbordone et al. 2005) 
and Pal~12 (Stetson et al. 1989; Brown et al. 1997;
Cohen 2004). The recent extensive and homogeneous analysis of relative ages of
Galactic GCs by Marin-Franch et al. (2009) identifies Pal 1, Pal 12 and Ter 7   
as the three youngest clusters of their whole sample.
In conclusion, the diagrams in the left panels of Fig.~\ref{f:rzfe}
are very effective in identifying as outliers three clusters that are (most
likely) of extra-galactic origin.

In the right panels of Fig. \ref{f:rzfe} we show the similar kind of plots
for the M31 GCs (metallicities from G09). Unfortunately, in the case of M31
we have at disposal only projected quantities (the projected galactocentric
distance $R_p$, and the projected distance from the major axis, a proxy for the
height above the disc), unavoidably blurring the
information contained in their de-projected  counterparts. Nevertheless, the
overall morphology of the distributions is quite similar to the MW case. In
particular there is just a bunch of metal-rich clusters having large $R_p$ and
$Y$, including B407. 

\begin{figure} 
\centering \includegraphics[width=8.cm]{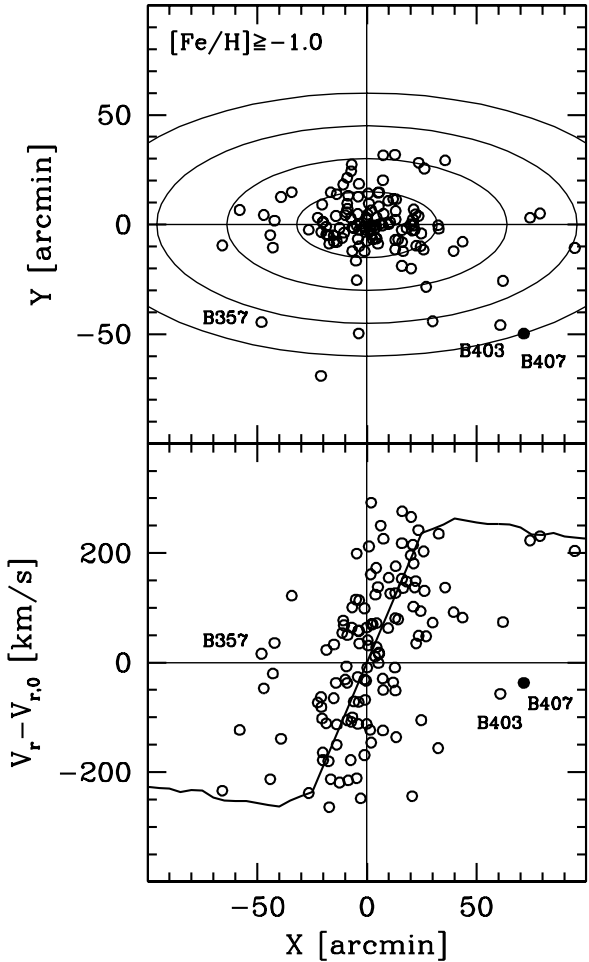}
\caption{Upper panel: X,Y distribution of M31 GCs having [Fe/H]$\ge -1.0$. The
ellipses overplotted have major axes of $30\arcmin$, $60\arcmin$,  $90\arcmin$
and $120\arcmin$, respectively, and have the same axis-ratio and orientation as
the disc of M31. Lower panel: Radial velocity as a function of major-axis
distance for the same M31 GCs as above. The line is the HI rotation curve of the
galaxy from Carignan et al. (2006). We have labelled only the clusters, among
those labelled in Fig.~\ref{f:rzfe}, that do not follow the general
rotation pattern. } 
\label{f:velpos} 
\end{figure}

To see if the anomaly in the position of these clusters can be traced also 
in their kinematics, in Fig. \ref{f:velpos} we plot the projected 
position of metal-rich ([Fe/H]$\ge -1.0$) clusters in the plane of the sky
(upper panel), and their M31-centric radial velocity as a function of their
distance along the major axis (assuming V$_{r,0}=-301$ km/s as the systemic
velocity of M31, Van den Bergh 2000). It is well known that, at odds with the MW case, the bulk of
M31 GCs participate to the rotation pattern of the galaxy disc, as traced by
the HI rotation curve, and the correlation is tighter for metal-rich clusters 
(P02;  Lee et al. 2008; G09, and references therein).
Among the clusters labelled in Fig.~\ref{f:rzfe} as having an anomalous position
for their metallicity, three have velocities in stark contrast with the rotation
pattern shared by the bulk of the metal-rich GCs: B357, B403 and B407. In
particular, the latter two clusters lie within a projected distance of 3 kpc
from each other, and have velocities differing by $\simeq 20$ km/s.
It is tempting  to suggest that the two clusters are (were) physically
associated to a common structure, having a different origin from the bulk of the
other clusters. Recent extensive surveys have revealed that the halo and the
outer disc of M31 host a wealth of sub-structures, generally
believed to be the relics of past accretion events that contributed to the
build-up of the galaxy (Ibata et al. 2001b, 2005, 2007; 
 Ferguson et al. 2005).

Indeed, the CMD of B407 presented here has been obtained from an image of the
set used by Richardson et al. (2008) to study the field stellar population within the
main sub-structures identified in Ferguson et al. (2005) and Ibata et al.
(2007). In particular, the cluster is at the edge of an ACS image sampling the
so called NE Shelf, a thin stream of stars looping over the North-Eastern edge
of the M31 disc. Richardson et al. find that the structure is mainly composed by
stars with metallicity similar to what we find for B407, also very similar
to the population found in the largest structure identified by Ferguson et al.
(2005)  and Ibata et al. (2007), i.e. the ''Giant Stream''.
Ibata et al. (2005) studied the kinematics of stars in a field of the NE Shelf
not far from B407/B403. They find a bimodal $V_{r}$ distribution with a major peak 
at the characteristic velocity of the M31 disc at this position ($V_{r} \sim  -200$km/s),
and a secondary peak at $V_{r} \sim  -350$km/s. We note that the velocities of the 
considered clusters match pretty well the secondary peak ($V_{r} = -338, -358 $km/s
for B407, B403 respectively), supporting the hypothesis of physical association with
the component of field stars that do not follows the kinematics of the disc.
\footnote{Another cluster, B401,
having very similar position and velocity ($X=56.99, Y=-32.30, V_{r}=-333$km/s), was
not plotted in  Fig. \ref{f:velpos} because of its very low metallicity ([Fe/H]=$-2.03$).}. 

The case described here opens a new window for the research of substructures in
M31, as it shows that it may be possible to use GCs to trace (and easily study
the kinematics) of at least some of the relics of past accretion events
(see also Lee et al. 2008).
It also supports the idea that the ingestion of GCs from accreting dwarf
galaxies may provide a significant contribution to the assembly of the 
globular cluster systems of giant galaxies, as already shown in the case of the
Milky Way (Bellazzini et al. 2003b)

\section{Summary and conclusions}

We have analysed 63 objects listed in the RBC  for which HST/ACS images were
publicly available in the HST Archive. We have confirmed or revised their
classification based on the inspection of these images and we were able to obtain
meaningful CMD for 11 likely old GCs and 6 young bright clusters.

We estimated distance, reddening, and metallicity for the eleven old GCs, by
comparing the field-decontaminated CMD of the clusters with a grid of ridge
lines of well-studied template clusters of the Milky Way. Our reddening and
metallicity estimates are, in general, in satisfactory agreement with previous
independent measures.  As reported in Table \ref{t:parme}, we have also
determined for each cluster an estimate of the magnitude level of the HB
measured on the HB ridge line of the  reference GGC that best fits the observed
CMD, with a typical error of $\pm$0.15 mag. 

One of the clusters of our sample (B407) is identified as a possible member of
a large sub-structure recently found in the halo of M31, and interpreted as
a relic of past (minor) merging episodes (NE Shelf; Ibata et al. 2001b, 2005,
2007; Richardson et al. 2008). The cluster has a metallicity that is much higher
than the bulk of M31 clusters residing at the same (large) distance from the M31
centre/major axis, and its kinematics is very different from the large
majority of M31 GCs having similar metallicity. The GC B403 (not included in our
sample) is found to share the same properties and is also indicated as a
possible member of the NE Shelf.

We estimated the age also for six candidate young clusters,  by comparing their
observed CMD with theoretical isochrones.  We confirm that all of them are
younger than 1 Gyr, in good agreement with previous studies.

With the present analysis the total number of M31 confirmed GCs with published
reliable optical CMDs increases from 35 to 44 for the old
globulars, and from 7 to 11 for the young bright ones (BLCCs).    The photometric
catalogues of the clusters studied here will be made publicly available through
a dedicated web page\footnote{\tt http://www.bo.astro.it/M31/hstcatalog}.

\begin{acknowledgements}
We are grateful to P. Barmby for providing her unpublished list of reddening 
estimates for M31 GCs. 
We thank the referee, M. Rich, for useful comments and suggestions.
This research was supported by ASI through the grant I/023/05/0. 
S.P., M.B. and C.C. acknowledge the financial support by INAF under the 
PRIN 2007 grant CRA 1.06.10.04.
\end{acknowledgements}


\end{document}